\documentclass[nonacm, acmsmall, screen, authorversion]{acmart}
\acmConference{arXiv Preprint}{August}{2021}
\graphicspath{{arXiv-figures/}}

\usepackage{amsmath, amsthm}
\usepackage{subcaption}
\usepackage{booktabs}
\usepackage{microtype}

\let\vec\mathbf
\newcommand*{\tran}{^{\mkern-1.5mu\mathsf{T}}}

\begin{document}

\title
[Neuromuscular Control of the Face-Head-Neck Biomechanical Complex With Learning-Based Expression Transfer]
{Neuromuscular Control of the Face-Head-Neck Biomechanical Complex With Learning-Based\\ Expression Transfer From Images and Videos}

\author{Xiao S.~Zeng}
\authornote{Co-first author.}
\email{stevennz@cs.ucla.edu}
\orcid{0000-0002-4337-3398}
\affiliation{%
  \institution{University of California, Los Angeles}
  \streetaddress{UCLA Computer Science Department, 404 Westwood Plaza}
  \city{Los Angeles}
  \state{California}
  \country{USA}
  \postcode{90095}
}

\author{Surya Dwarakanath}
\authornotemark[1]
\email{suryadwar@g.ucla.edu}
\affiliation{%
  \institution{University of California, Los Angeles}
  \streetaddress{UCLA Computer Science Department, 404 Westwood Plaza}
  \city{Los Angeles}
  \state{California}
  \country{USA}
  \postcode{90095}
}

\author{Wuyue Lu}
\email{luwuyue@cs.ucla.edu}
\affiliation{%
  \institution{University of California, Los Angeles}
  \streetaddress{UCLA Computer Science Department, 404 Westwood Plaza}
  \city{Los Angeles}
  \state{California}
  \country{USA}
  \postcode{90095}
}

\author{Masaki Nakada}
\email{masaki@neuralx.ai}
\affiliation{%
  \institution{NeuralX, Inc., Los Angeles}
  \city{Los Angeles}
  \state{California}
  \country{USA}
  \postcode{90024}
}

\author{Demetri Terzopoulos}
\email{dt@cs.ucla.edu}
\affiliation{%
  \institution{University of California, Los Angeles}
  \streetaddress{UCLA Computer Science Department, 404 Westwood Plaza}
  \city{Los Angeles}
  \state{California}
  \country{USA}
  \postcode{90095}
}

\renewcommand{\shortauthors}{X.~S.~Zeng, S.~Dwarakanath, W.~Lu, M.~Nakada, and D.~Terzopoulos}

\begin{abstract}
The transfer of facial expressions from people to 3D face models is a classic computer graphics problem. In this paper, we present a novel, learning-based approach to transferring facial expressions and head movements from images and videos to a biomechanical model of the face-head-neck complex. Leveraging the Facial Action Coding System (FACS) as an intermediate representation of the expression space, we train a deep neural network to take in FACS Action Units (AUs) and output suitable facial muscle and jaw activation signals for the musculoskeletal model. Through biomechanical simulation, the activations deform the facial soft tissues, thereby transferring the expression to the model. Our approach has advantages over previous approaches. First, the facial expressions are anatomically consistent as our biomechanical model emulates the relevant anatomy of the face, head, and neck. Second, by training the neural network using data generated from the biomechanical model itself, we eliminate the manual effort of data collection for expression transfer. The success of our approach is demonstrated through experiments involving the transfer onto our face-head-neck model of facial expressions and head poses from a range of facial images and videos.
\end{abstract}



\begin{teaserfigure} \centering
\subcaptionbox{Subject 1}{\includegraphics[height=0.215\linewidth]{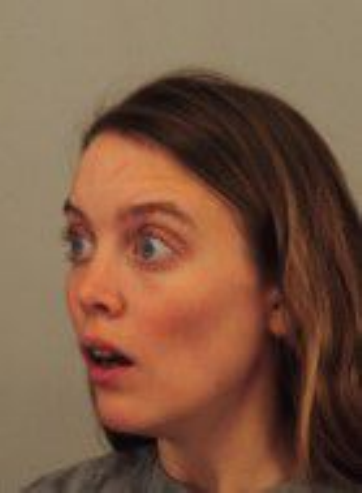}
\includegraphics[height=0.215\linewidth]{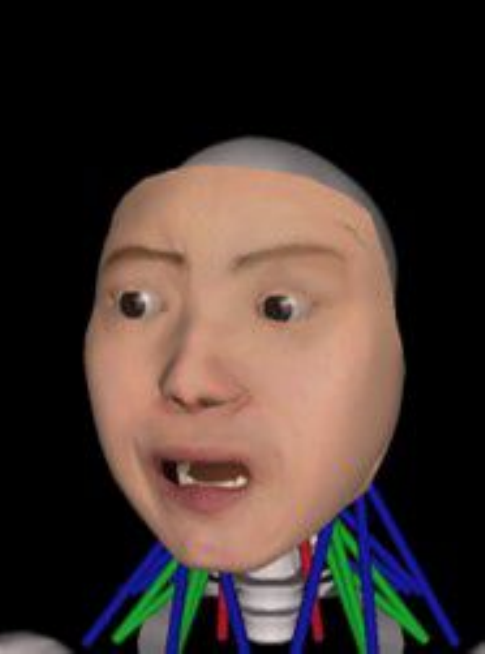}}
\hfill
\subcaptionbox{Subject 2}{\includegraphics[height=0.215\linewidth]{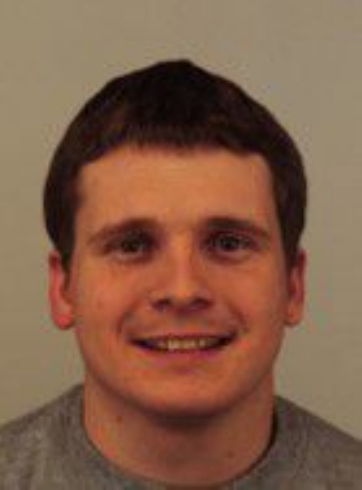}
\includegraphics[height=0.159\linewidth, height=0.215\linewidth]{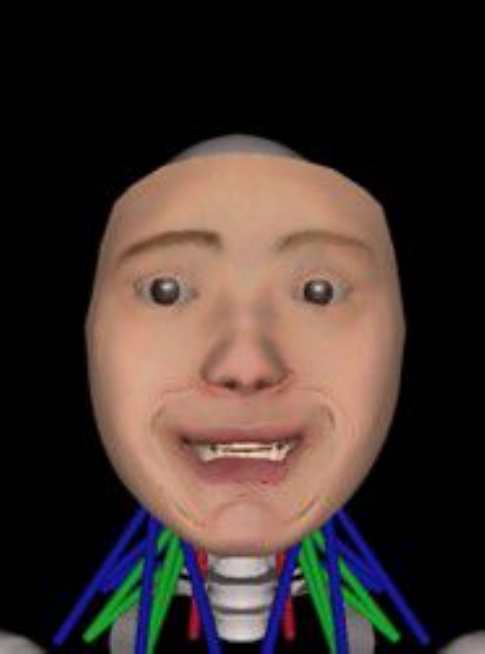}}
\hfill
\subcaptionbox{Subject 3}{\includegraphics[height=0.215\linewidth]{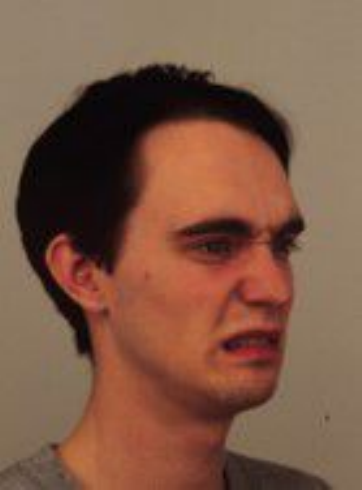}
\includegraphics[height=0.215\linewidth]{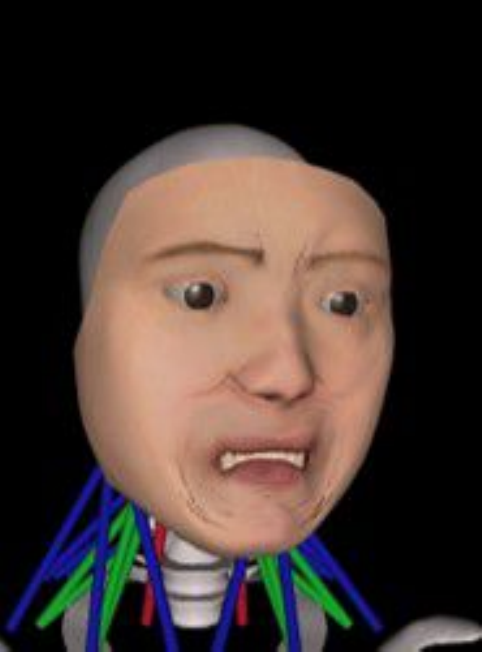}}
\caption{Our deep learning framework enables the transfer of facial
expressions and head poses from images or videos to a biomechanical
face-head-neck model.}
\label{fig:teaser}
\end{teaserfigure}

\maketitle

\section{Introduction}

The face, actuated by the muscles of facial expression, and head
movements, actuated by the cervical muscles, are a powerful mode of
nonverbal communication between humans. The simulation of the
face-head-neck musculoskeletal complex is of importance in
understanding how we convey thinking and feeling in fields from
affective science to 3D computer animation. Biomechanical human
musculoskeletal models realistically capture the anatomy and physics
underlying human motion generation. In particular, those pertaining to
the simulation of the human face have succeeded in convincingly
emulating facial expressiveness; however, they require significant
effort in parameter tuning to produce realistic results.

In this paper, we show how to endow a biomechanical musculoskeletal
model of the human face with the ability to produce facial expressions
via machine learning from real-world reference images and videos
(Figure~\ref{fig:teaser}). To this end, we introduce a deep
neural-network-based method for learning the representation of human
facial expressions through Ekman's Facial Action Coding System (FACS)
\cite{cohn2007observer} in the context of the muscle actuators that
drive the musculoskeletal face model augmented with a musculoskeletal
cervicocephalic (neck-head) system to animate head movement during
facial expression synthesis.

\begin{figure} \centering
\subcaptionbox{Framework\label{fig:framework}}
{\includegraphics[trim={30 0 0 50},clip, width=0.53\linewidth]{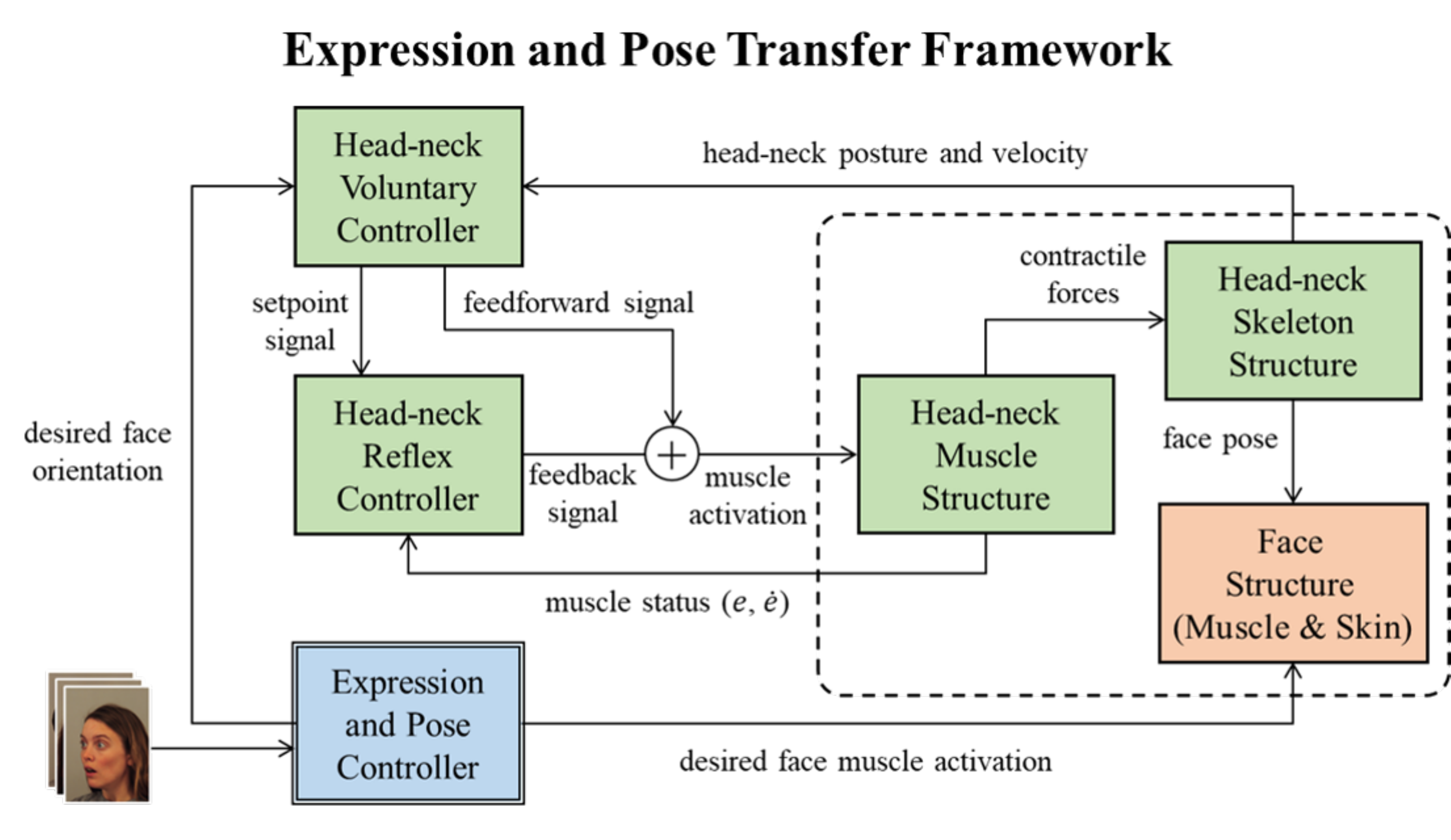}}
\hfill
\subcaptionbox{Controller}
{\includegraphics[width=0.44\linewidth, trim={0 20 20 70}, clip]{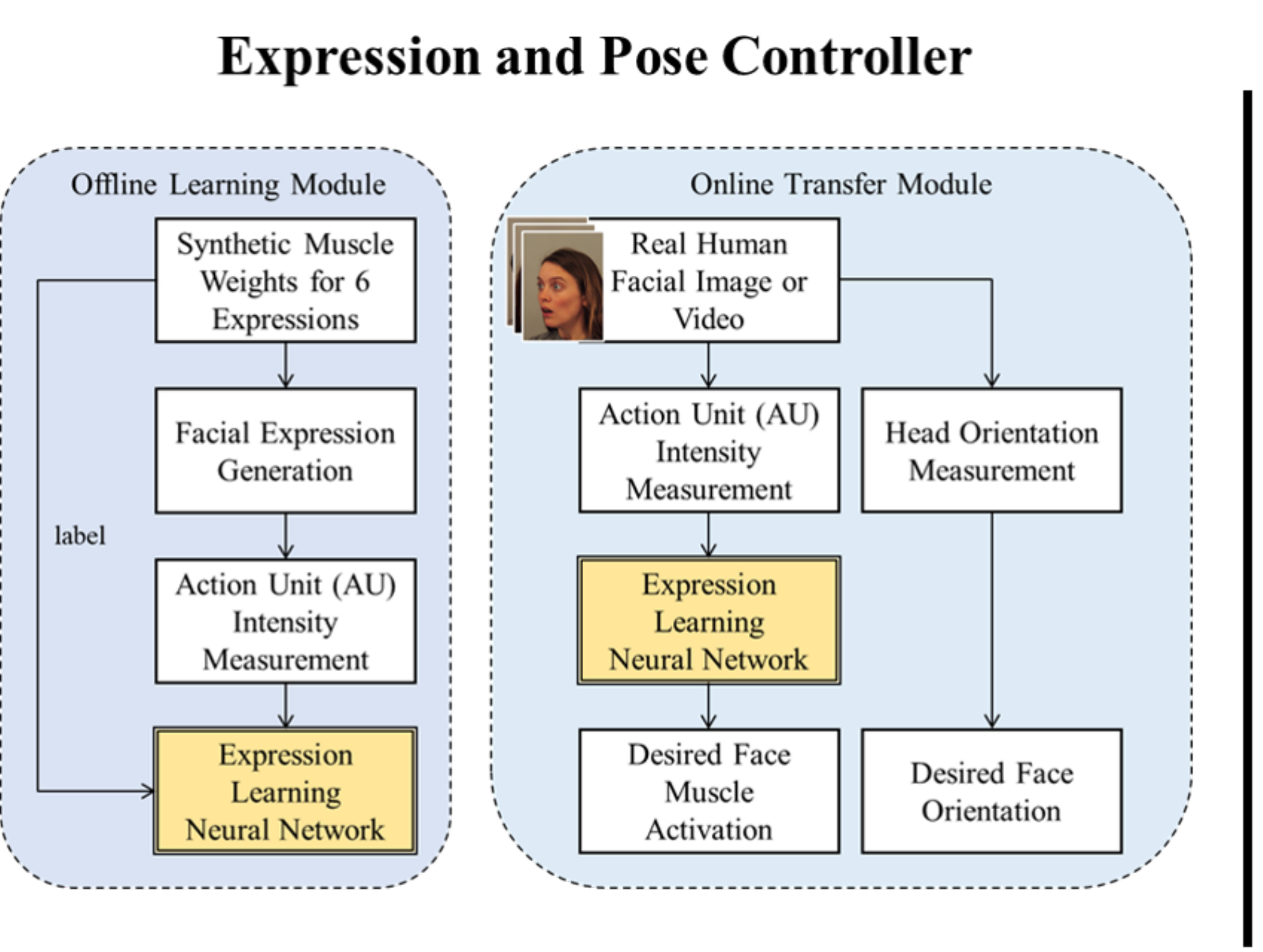}}
\caption{The structure of our expression and head pose transfer
framework (a) and the components of the facial expression and head
pose controller (b). The expression learning neural network (yellow)
is first trained offline. In a transfer task, an image or a video
sequence of a real face is fed into the online transfer module of the
expression and pose controller to output the desired facial muscle
activations and head orientation information. The muscle activations
are input to the biomechanical face model (orange) to perform the
corresponding expression and the head orientation is provided to the
head-neck biomechanical neuromuscular system (green) to produce the
desired head pose.}
\label{fig:structure_overview}
\end{figure}

The novelty of our framework lies in the following features:
\begin{enumerate}
\item
We propose the first biomechanical face-head-neck animation system
that is capable of learning to reproduce expressions and head
orientations through neuromuscular control.
\item
Our novel deep neuromuscular motor controller learns to map between
FACS Action Units (AUs) extracted from human facial images and videos
and the activations of the muscle actuators that drive the
biomechanical system.
\item
As a proof of concept, we demonstrate an automated processing pipeline
(Figure~\ref{fig:structure_overview}) for animating expressions and head
poses using an improved version of the physics-based neck-head-face
animation system developed by Lee and Terzopoulos~\cite{lee2006heads},
but which can potentially be applied to any muscle-driven model.
\end{enumerate}

\section{Related Work}

Muscle-based facial animation has been studied for decades
\citep{waters1987muscle}. A series of studies have endeavored to build
increasingly anatomically accurate musculoskeletal face models. Using
such models, \citet{terzopoulos1993analysis}
and \citet{sifakis2005automatic}, among others, have
addressed the automatic determination of facial muscle activations
from videos, but these methods require nonrigid facial motion trackers
or motion capture markers. In this context, the work by 
\citet{ishikawa1998, ishikawa20003d} is of particular interest
in that it employed three trained neural networks to transduce from 
the positions of facial markers to the activations of facial muscles. 
We present a markerless, real-time
biomechanical face-head-neck simulation system that can automatically
learn muscle activation parameters for a variety of expressions from
reference images and videos. Most relevant to our work in the present
paper is an existing biomechanical model of the cervicocephalic
complex \citep{lee2006heads}, which incorporates an improved version
of the face model developed by \citet{lee1995realistic}.

Many facial animation researchers have made use of the well-known FACS
\citep{ekman1978manual}, a quantitative phenomenological abstraction
that decomposes facial expressions into the intensity levels of Action
Units (AUs), each of which corresponds to the actions of one or more
facial muscles. An advantage of using the FACS is that it encodes
anatomical and psychological semantics \citep{cohn2007observer}.

Expression transfer or retargeting is a trending topic and recent
approaches often use FACS-based blendshapes as the basic parametric
representation
\cite{weise2011realtime,thies2016face2face,zhang2020facial}. Others
have used techniques such as interactive mesh deformation control
\citep{xu2014controllable} and neural-network-based perceptual models
\citep{aneja2016modeling} to represent expressions in blendshapes.
However, transferring expressions using a musculoskeletal system is
more natural since facial actions are the result of coordinated muscle
contractions inducing soft tissue deformation. Despite the existing
literature, there remains a deficiency of work in transferring
expressions to musculoskeletal facial systems.

\section{Musculoskeletal Model}
\label{sec:Musculoskeletal}

Our real-time musculoskeletal model is based on the one of 
\citet{lee2006heads}, but both the underlying
face-head-neck control system and the facial expression system are
significantly improved. Its overall architecture is controlled in a
hierarchical manner, as illustrated in Figure~\ref{fig:framework}.
Specifically, the skeletal structure is an articulated multibody
dynamics system, with bones and joints consistent with human anatomy.
The skeletal system is driven by a Hill-type muscle actuator model.

The biomechanical face component consists of a facial soft tissue
model comprising epidermis, dermal-fatty, fascia, and muscle layers
supported by an underlying skull, which is constructed based on the
work of \citet{lee1995realistic}. The soft tissue is a
deformable model assembled from discrete uniaxial finite elements,
which simulates dynamic facial deformation in an anatomically
consistent yet simplified way compared to the models described in
\citep{sifakis2005automatic}, \citep{wu2014generating}, and
\citep{cong2015fully}, thus maintaining a low computational cost that
affords real-time simulation performance on readily available
hardware.

There are 26 pairs of primary facial muscles embedded in the
biomechanical face, including the frontalis, corrugator, levator
labii, orbicularis oculi, mentalis, and orbicularis oris muscle
groups. The contractions of these muscles apply forces to the facial
soft tissue layers, inducing deformations that produce meaningful
facial expressions. We have augmented the expressive details, such as
wrinkles on the face model, by applying multiple levels of subdivision
to increase significantly the number of surface nodes that can be
activated by muscle forces, and applied a high resolution texture map
to the surface mesh for a natural appearance.

Appendix~\ref{app:details} explains the biomechanical components of
our model in greater detail.

\subsection{Control}

To control the face-head-neck system, our novel neural network-based
expression and pose controller generates facial muscle activations
that produce recognizable expressions. It concurrently outputs head
pose estimates to the head-neck musculoskeletal complex, where
voluntary and reflex neuromuscular control layers generate muscle
activation signals to achieve the desired head orientations.

The higher-level voluntary controller receives the current head-neck
posture and velocity information, as well as the desired adjustment of
the posture, and generates a feedforward muscle activation signal and
a setpoint control signal. The latter, which encodes the desired
muscle strains and strain rates, is input to the reflex controller
that then generates a feedback muscle activation signal and adds it to
the feedforward signal generated by the voluntary controller. As a
result, each cervical muscle receives an activation signal $a$ and,
through simulation, generates a contractile muscle force accordingly.
Together with the external environmental forces and gravity, the whole
system is simulated through time and rendered as a physics-based
animation. The voluntary controller runs at 25Hz (in simulation time),
while the reflex controller runs at 4KHz along with the physical
simulation time steps.

\section{Expression Learning}
\label{sec:Expression}

We next explain our machine learning approach of using a
deep neural network to transfer facial expressions to our
biomechanical face model. We leverage the FACS and synthesize the
muscle activations for the model using the trained deep neural network. The
following sections describe the architecture of the network, the
pipeline for using the biomechanical face model to generate training
data, and the process of training the neural network.

\subsection{Network Architecture}

\begin{figure} \centering
\subcaptionbox{Neural network architecture \label{fig:nn}}
{\includegraphics[width=0.48\linewidth]{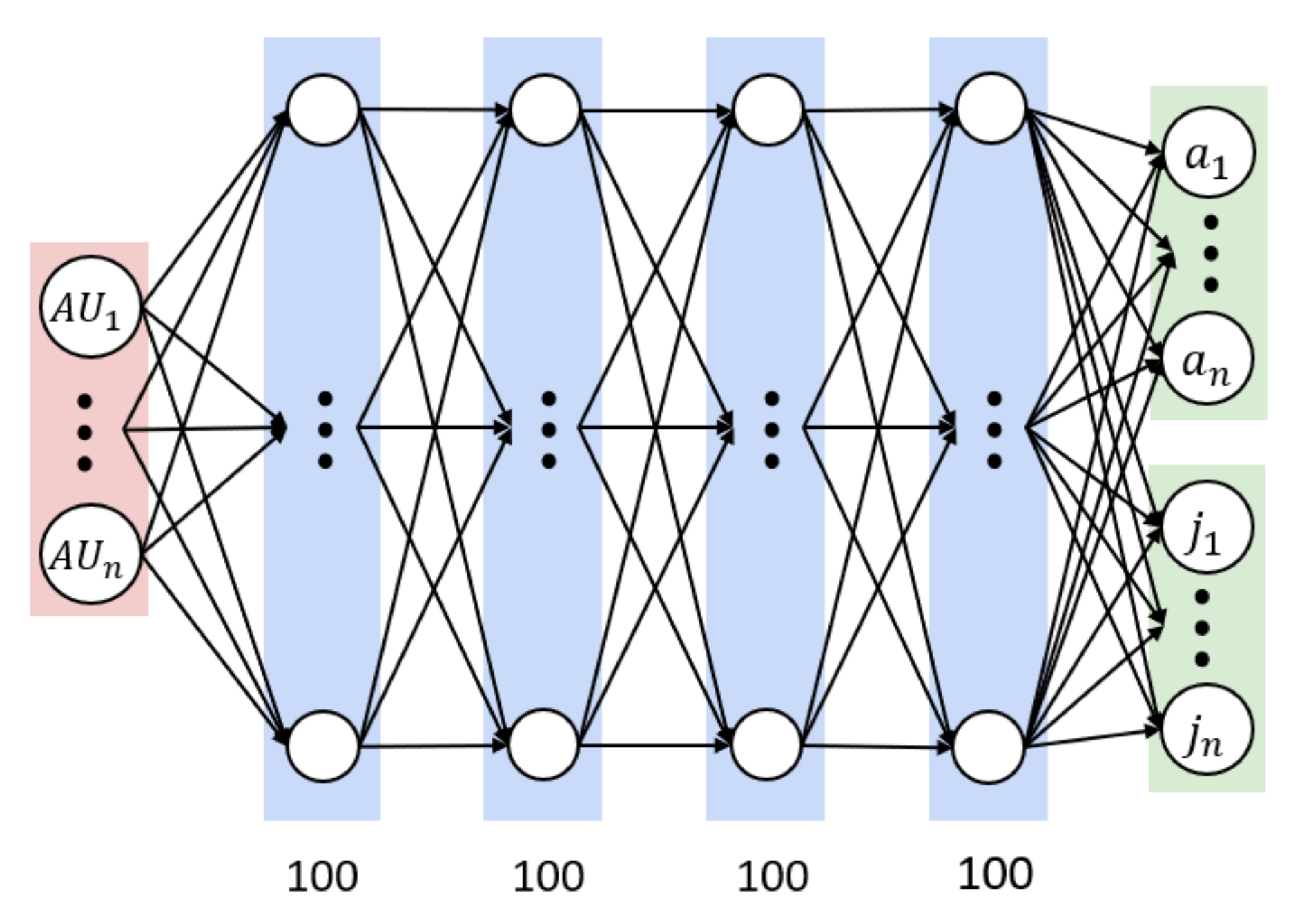}}
\hfill
\subcaptionbox{Convergence of the neural network \label{fig:loss}}
{\includegraphics[width=0.48\linewidth, trim={0 0 40 40},clip]{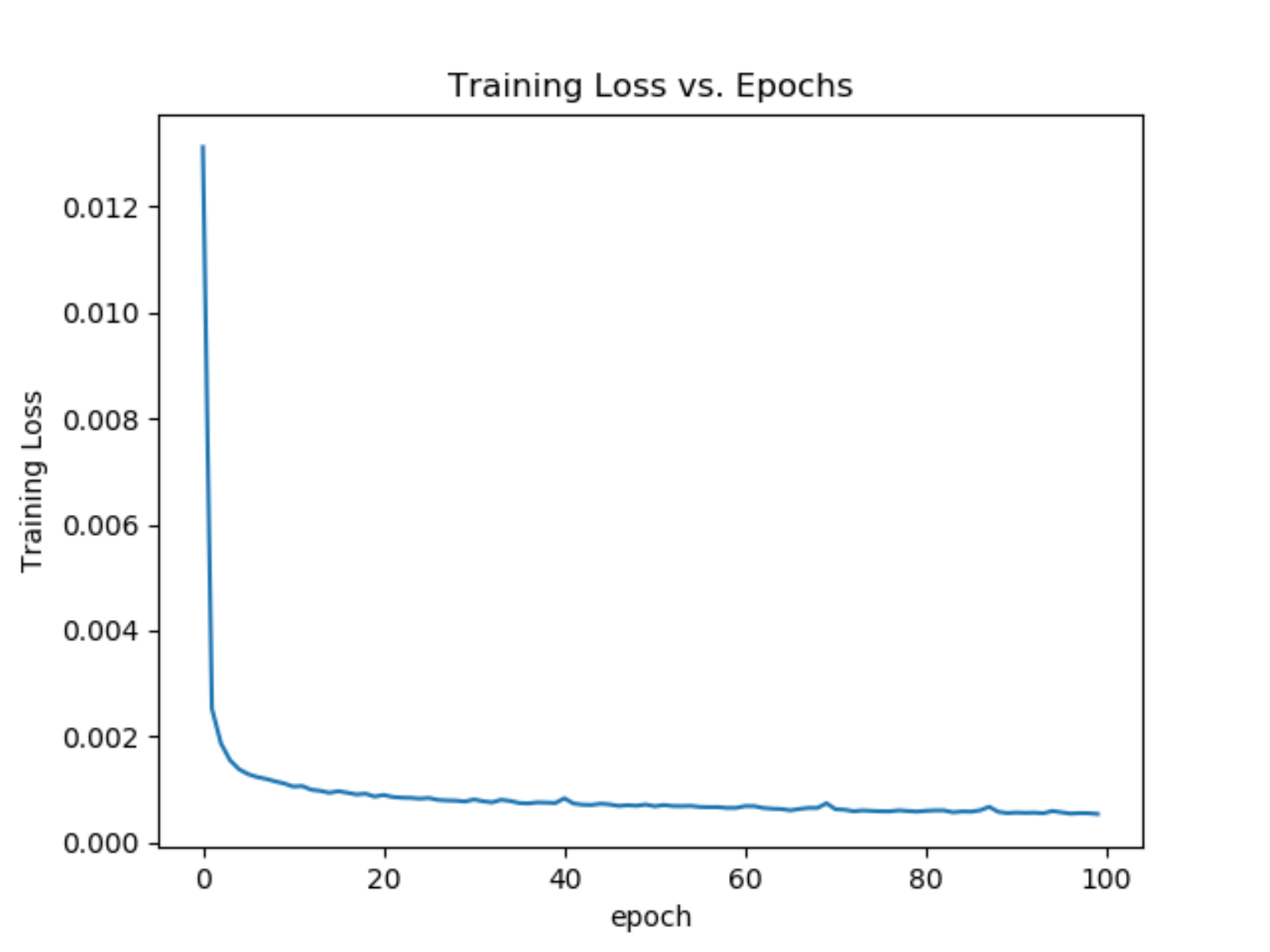}}
\caption{The expression learning neural network.}
\label{fig:expr_NN}
\end{figure}

The function of the neural network is to generate activation signals
for the muscles of the biomechanical face model such that they
generate contractile forces that deform the synthetic facial tissues
to produce facial expressions. We employ a fully connected deep neural
network architecture for this purpose (Figure~\ref{fig:nn}). The input
layer consists of a total of 17 neurons representing the important AUs
that are involved in the majority of facial movements, each neuron
corresponding to a single normalized AU. We include 4 hidden layers
with 100 neurons in each layer and ReLU activation.\footnote{This
choice was based on observed Mean Squared Errors (MSE) on test data of
networks with different numbers of hidden layers and different numbers
of neurons per hidden layer.} The output layer consists of 56 neurons,
52 of which encode the activations $a_i$, with $1 \leq i \leq 52$, for
each of the 26 pairs of facial muscles, and the remaining 4 encode the
jaw rotation, jaw slide, jaw twist, and an auxiliary value. Given its
architecture, the network has a total of 37,300 weights. It is
implemented in Keras with a TensorFlow backend.

\subsection{Training Data Generation}

The training data generation process is divided into two steps:
\begin{enumerate}
\item generation of muscle activations, and
\item generation of AUs for the corresponding muscle activations.
\end{enumerate}

Each basic expression requires a combination of muscles to be
activated. Given $n$ muscles, we define $W_e$ as a set of weights
$w_{i}$, with $1\leq i \leq n$, which determine the effect each muscle
will have on an expression $e$. The activation for each muscle $a_{i}$
for an expression is then defined as $a_{i} = w_{i}s$, where $w_{i}
\in W_e$ and $s$ is a scale term. For a single expression, we
determine the weights in a set $W_e$ manually by visually analyzing
the facial expressions formed by the face model. We repeat the process
for all the basic expressions, namely (1) Joy, (2) Sadness, (3) Anger,
(4) Fear, (5) Disgust, and (6) Surprise to generate the sets
$W_\text{Joy}$, $W_\text{Sadness}$, $W_\text{Anger}$, $W_\text{Fear}$,
$W_\text{Disgust}$ and $W_\text{Surprise}$, respectively.

We then sample the value of $s$ randomly in the range $[0.0, 1.0]$ to
generate all the muscle activations $a_{i}$. We also assign a random
value between $0$ and $1$ for the jaw rotation. For the purpose of our
experiments, we maintain the jaw twist and jaw slide at a value of
$0.5$. We further form a set $A$ consisting of all the muscle
activations $a_{i}$, where $1\leq i \leq n$, and the jaw activations
including jaw rotation, jaw twist and jaw slide. We iterate the above
process, generating the set $A$ by repeatedly sampling the value of
$s$ for a single expression. We finally extend this to all basic
expressions and obtain multiple sets $A_\text{Joy}$,
$A_\text{Sadness}$, $A_\text{Anger}$, $A_\text{Fear}$,
$A_\text{Disgust}$ and $A_\text{Surprise}$ for each expression.

We leverage the functionality of OpenFace 2.0
\citep{baltrusaitis2018openface} for facial expression recognition and
head pose estimation. OpenFace is a computer vision tool capable of
facial landmark detection, head pose estimation, facial AU
recognition, and eye-gaze estimation. OpenFace employs Convolutional
Experts Constrained Local Models (CE-CLMs) for facial landmark
detection and tracking. CE-CLMs use a 3D representation of the
detected facial landmarks by which OpenFace estimates the head pose.
Eye gaze estimation is done using a Constrained Local Neural Fields
(CLNF) landmark detector. The final task, facial expression
recognition, is performed using Support Vector Machines and Support
Vector Regression.

We use a single set $A$ formed by the above described procedure to
activate the muscles and jaw of the biomechanical face model. We then
render the model to form an image. The image is input to OpenFace,
which performs facial expression recognition and head pose estimation,
outputting the estimated AUs and head orientation associated with the
input image. We repeat this process for each set $A$ formed (as
described previously) to obtain the corresponding AUs and head
orientations.

Thus, we synthesize a large quantity of training data pairs each
consisting of (i) muscle and jaw activations $A$ and (ii) the
associated AUs and head orientations.

\subsection{Network Training}

We use the aforementioned data to train our deep neural network to
input AUs and output corresponding muscle and jaw activations.

The AUs from each training pair, generated as described in the
previous section, are passed to the network as input. We then compare
the corresponding muscle and jaw activations; i.e., the ground truth
compared to the predictions of muscle and jaw activations given by the
neural network. We use a Mean Square Error (MSE) training loss between
the predictions and the ground truth, which is backpropagated to
update the weights, thus training the neural network.

We normalize the intensity values of each AU class across all pairs to
remove the disparity of intensity values between synthetic faces and
real faces. We use a total of 6,000 pairs, with about 1,000 pairs for
each basic expression.

To train the neural network, we use the ADAM stochastic gradient
descent optimizer with an MSE loss, a batch size of 32, and a learning
rate of 0.01. We train the network in a total of 100 epochs, running
on an NVIDIA GeForce GTX 1650 GPU installed on a Windows 10 machine
with a 2.6GHz Intel Core i7-9750H CPU. Figure~\ref{fig:loss} shows the
convergence of the training error.

\subsection{Expression Transfer Pipeline}

To transfer real facial expressions on the fly, we use a pipeline
similar to the offline training module, again leveraging OpenFace for
facial expression recognition and head pose estimation. We input an
image of an expressive face into OpenFace to obtain all of the
corresponding AUs and head orientations. The AUs are then passed into
the trained neural network which outputs predictions of the muscle and
jaw activations, driving the biomechanical face to deform the muscles
and transfer the expressions onto it.

We transfer both image and video inputs. Each frame in a video is
processed independently and a resulting video is created using the
transferred frames. The intensity values for each AU class are
normalized across all the images or frames as in the case of the
training pipeline.

\section{Experiments and Results}

We next present the results of transferring facial expressions from
the wild using our trained neural network. We evaluate our expression
transfer pipeline on different expressions while using a variation of
AUs and muscles in the biomechanical face model.

\subsection{Facial Expression Datasets}

Several facial expression datasets are available online. The datasets
that we used in our experiments are as follows:

\textit{Karolinska Directed Emotional Faces (KDEF)} \citep{Calvo2008}.
The KDEF dataset consists of 4,900 pictures of human facial
expressions. It covers 70 subjects (35 female and 35 male) enacting
all the basic facial expressions, namely Neutral, Joy, Sadness, Anger,
Fear, Disgust, and Surprise. Each expression performed by the subject
is imaged from multiple directions. We use the dataset to transfer
facial expressions onto the biomechanical face model and visually
analyze the performance of our trained neural network in this paper.

\textit{Cohn Kanade Dataset (CK)} and \textit{Extended Cohn Kanade
Dataset (CK+)} \citep{840611,5543262}. The CK and the CK+ dataset
combined consist of 593 video sequences of 123 subjects. Each sequence
consists of images from a neutral expression (first frame) to a peak
expression (last frame). The peak expressions are FACS coded for AUs.
We use the sequences in the CK+ dataset to transfer videos of
expression transitions onto the biomechanical face.

\textit{Ryerson Audio-Visual Database of Emotional Speech and Song
(RAVDESS)} \citep{10.1371/journal.pone.0196391}. The original RAVDESS
dataset consists of 24 actors vocalizing two lexically-matched
statements. An extension of the dataset, named \textit{RAVDESS Facial
Landmark Tracking}, contains tracked facial landmark movements from
the original RAVDESS datasets \citep{swanson2019}. Each actor performs
60 speech trials and about 44 song trials. This yields a total of
2,452 video files for the complete dataset. We leverage this dataset
to test the transfer of actor faces in speech videos onto the
biomechanical face.

\subsection{Action Units and Muscle Activations}

There exist a total of 30 Action Units (AUs) corresponding to facial
expressions. OpenFace provides occurrence predictions for 18 out the
30 AUs and measures intensity values for 17 out of the 30 AUs. We
consider the 17 AUs for which the intensity values are present as the
super-set of the AUs for our use case.

Due to the correlation between the AUs and muscles in the face, there
also exists a correlation between a basic facial expression and the
AUs activated by it. Our initial experiments focused on training the
neural network for each expression in an isolated manner. The neural
network was trained to output muscle activation for muscles
corresponding to a single expression using AUs which pertained to the
same expression. In further trials, we observed that usage of all 17
AUs and all facial muscles improved the performance and the
scalability of the expression transfer pipeline.

Table.~\ref{table:individual_training} provides a comparison between
training a single neural network for all expressions and training
individual neural networks for each expression, where each network is
trained only from the AUs and muscles relevant to its expression
(using the mappings of expressions to AUs and AUs to muscles presented
in \cite{DuE1454}, \cite{clark2020facial} and
\cite{10.3389/fpsyg.2019.00508}). We observe better performance when
training a single neural network for all the expressions, suggesting
that AUs not directly relevant to an expression also play a role in
expression transfer.

\begin{table} \centering
\caption{Comparison of the Mean Squared Errors for training individual
neural networks for each expression and for training a single neural
network for all expressions.}
\label{table:individual_training}
\begin{tabular}{c c c c c c c} 
\toprule
$\text{MSE}_{\text{Joy}}$ & $\text{MSE}_{\text{Sadness}}$ &
$\text{MSE}_{\text{Fear}}$ & $\text{MSE}_{\text{Anger}}$ &
$\text{MSE}_{\text{Surprise}}$ & $\text{MSE}_{\text{Disgust}}$ &
$\text{MSE}_{\text{All}}$\\
\midrule 0.000591 & 0.002925 & 0.001611 & 0.009689 & 0.000266 &
0.002516 & 0.000729 \\
\bottomrule
\end{tabular}
\end{table}

\begin{figure} \centering
\subcaptionbox{Subject 1}
{\includegraphics[width=0.24\linewidth]{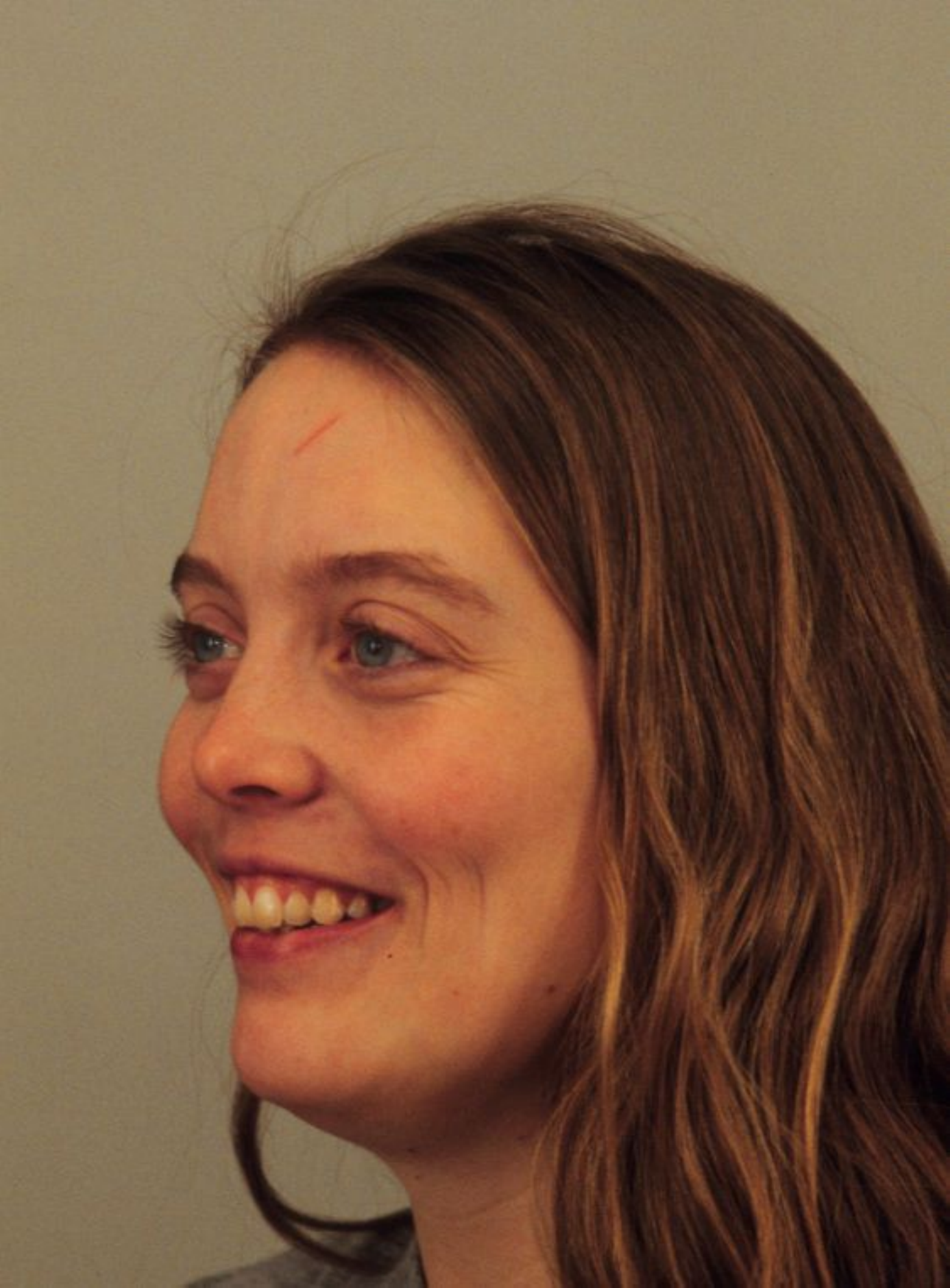} \includegraphics[width=0.24\linewidth]{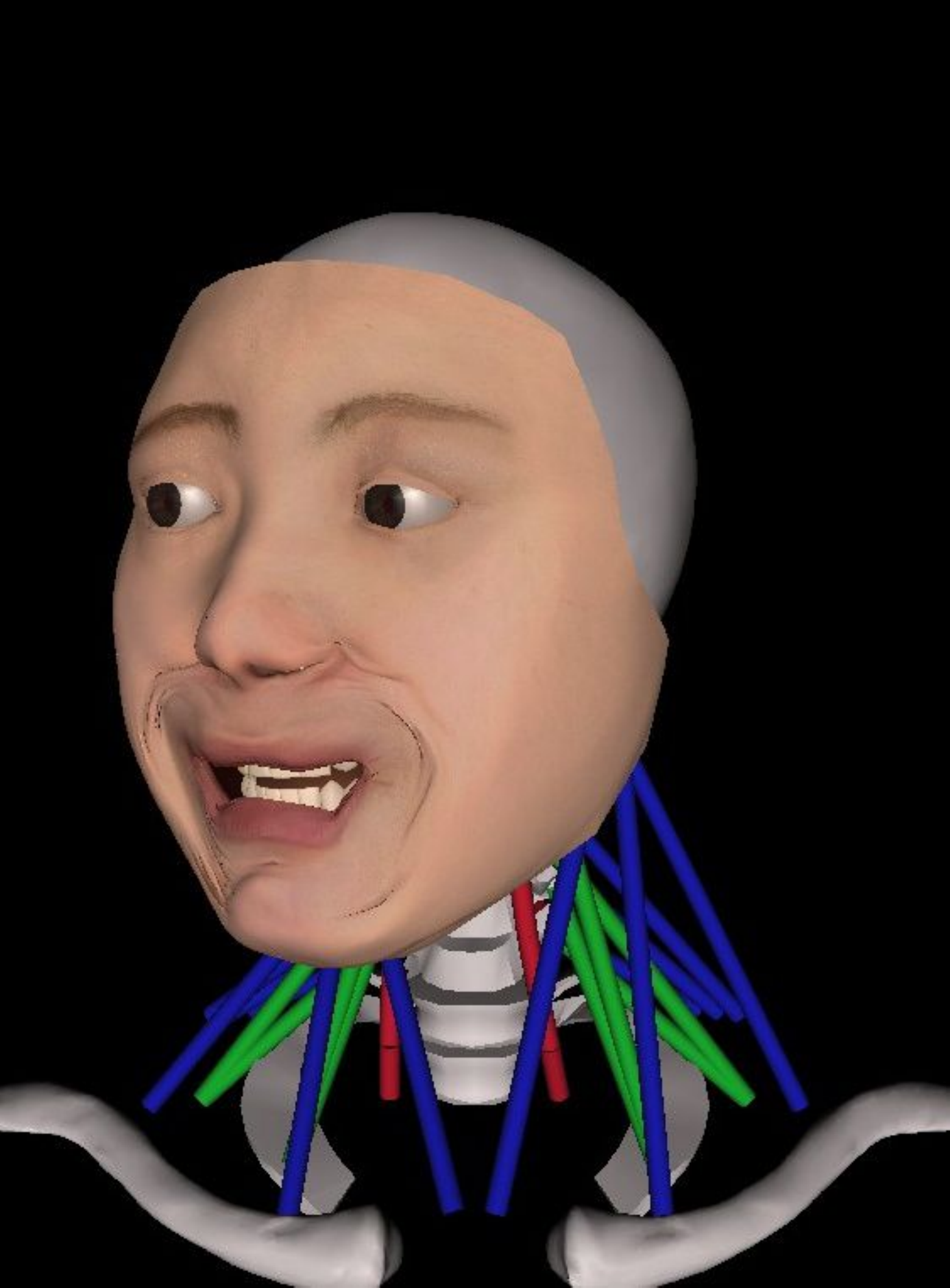}}
\hfill
\subcaptionbox{Subject 3}
{\includegraphics[width=0.24\linewidth]{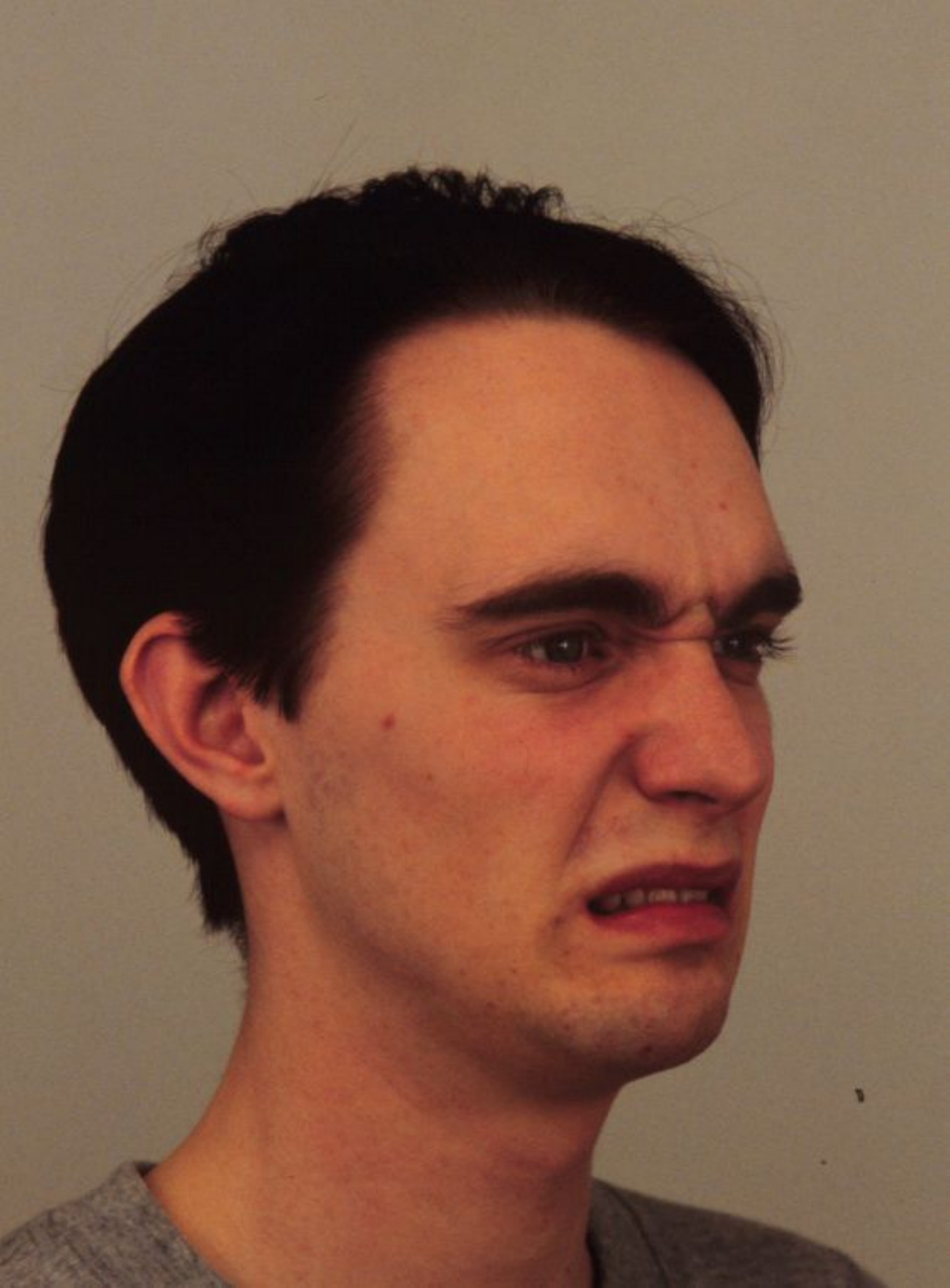}
\includegraphics[width=0.24\linewidth]{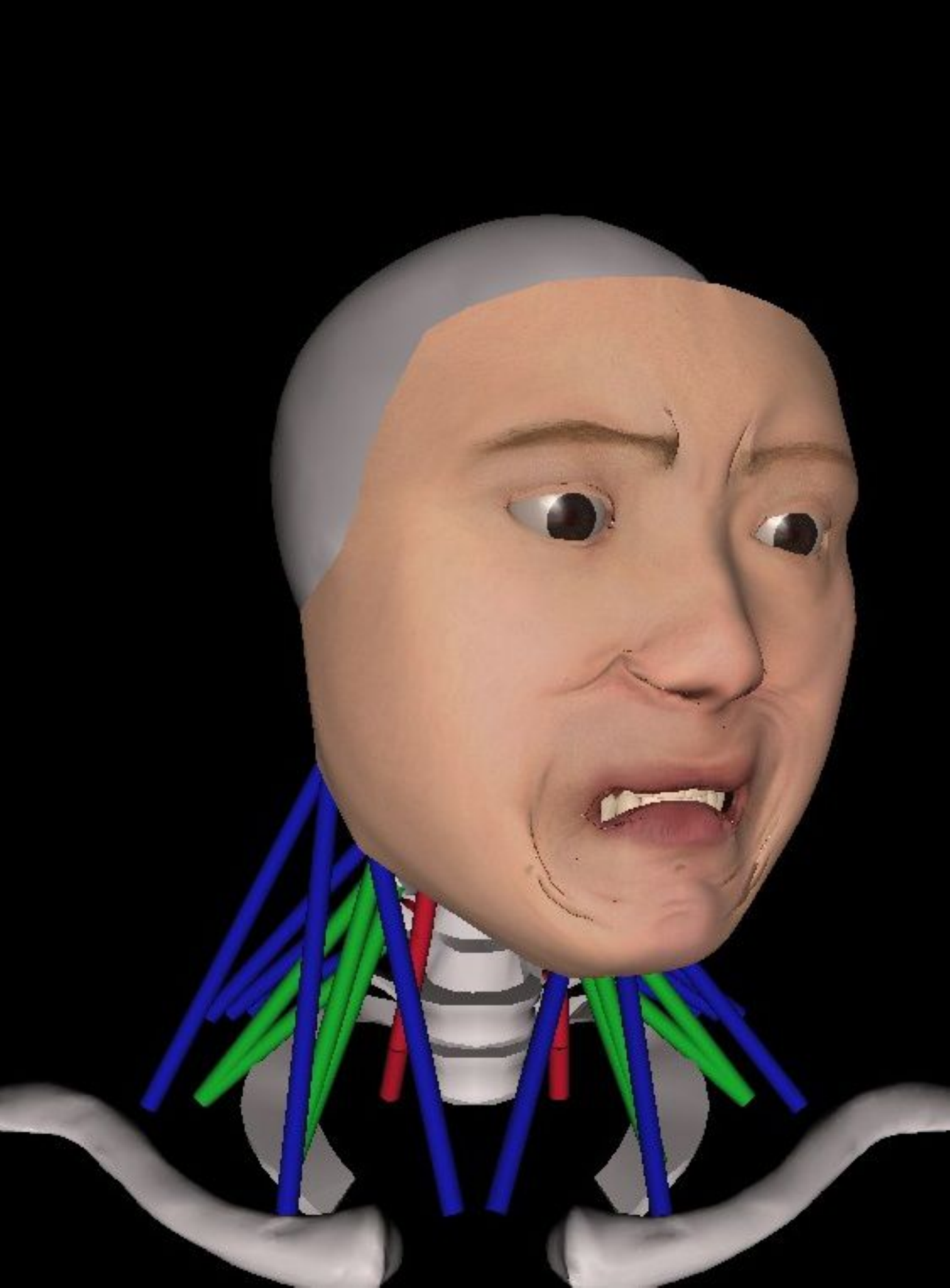}}
\caption{Transfer of head orientation along with facial expression.}
\label{fig:orientations}
\end{figure}

Due to limitations in the biomechanical model, the range of each AU
class differs from that of the real faces. Hence, we use normalization
to overcome the bias and better transfer real facial expressions onto
the biomechanical model. The expression intensities in transferred
expressions with normalization better represent the real faces than
those without normalization.

We choose to activate only jaw rotation so as to maintain the symmetry
of the expression for our use case. We observe, that without jaw
activations, expressions such as surprise are not well synthesized by
the biomechanical face model.

\subsection{Head Orientation}

Leveraging OpenFace for head pose estimation, we pass the estimated
orientation of the head into the trained neck controller to activate
the neck muscles. This in turn actuates the neck to adjust the head
orientation in accordance with the input image.
Figure~\ref{fig:orientations} presents sample transferred results
including head orientations from the KDEF dataset.

\begin{figure} \centering
\subcaptionbox{Subject 4}[\linewidth]{\includegraphics[width=0.158\linewidth]{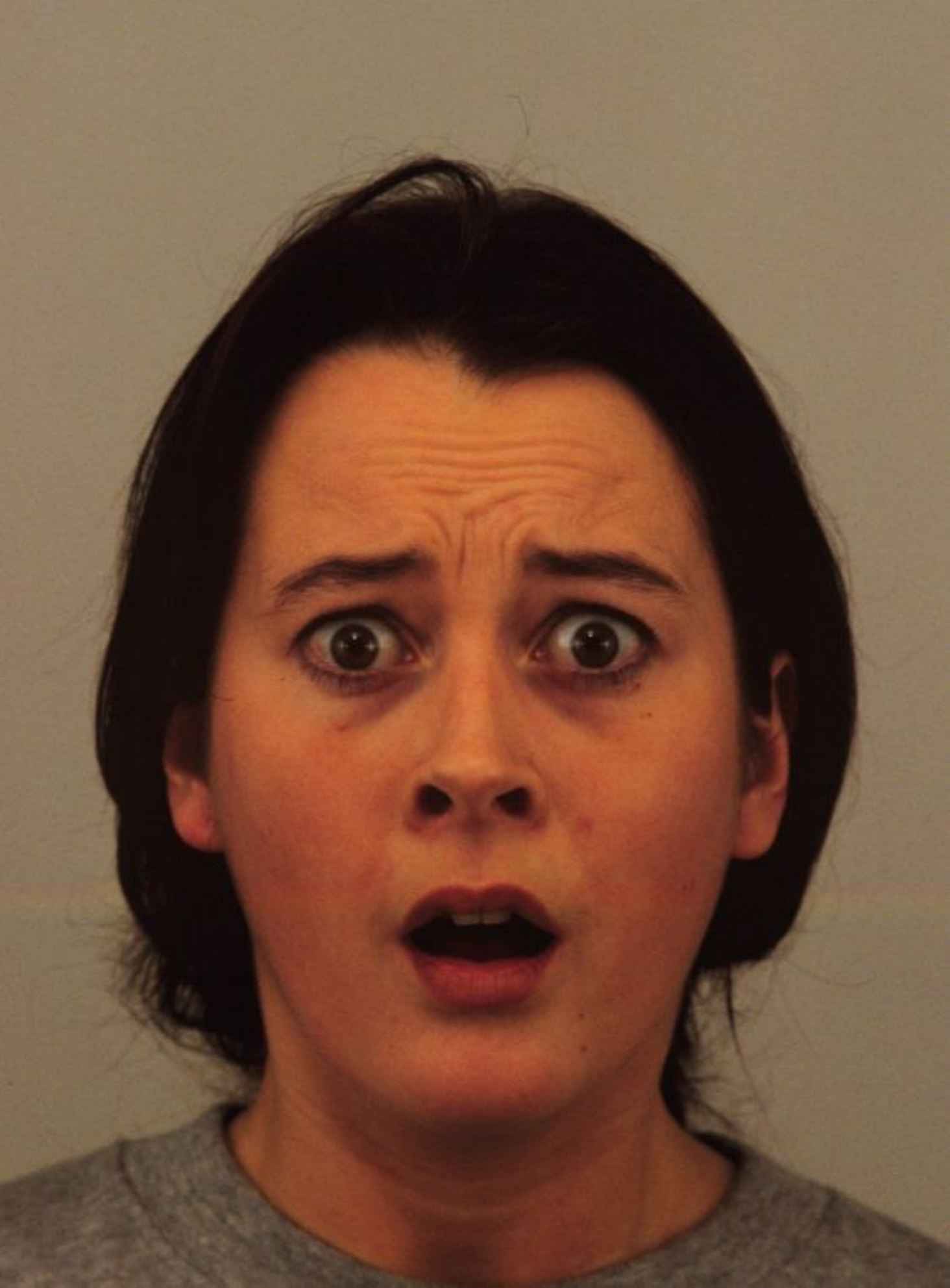}
\hfill
\includegraphics[width=0.158\linewidth]{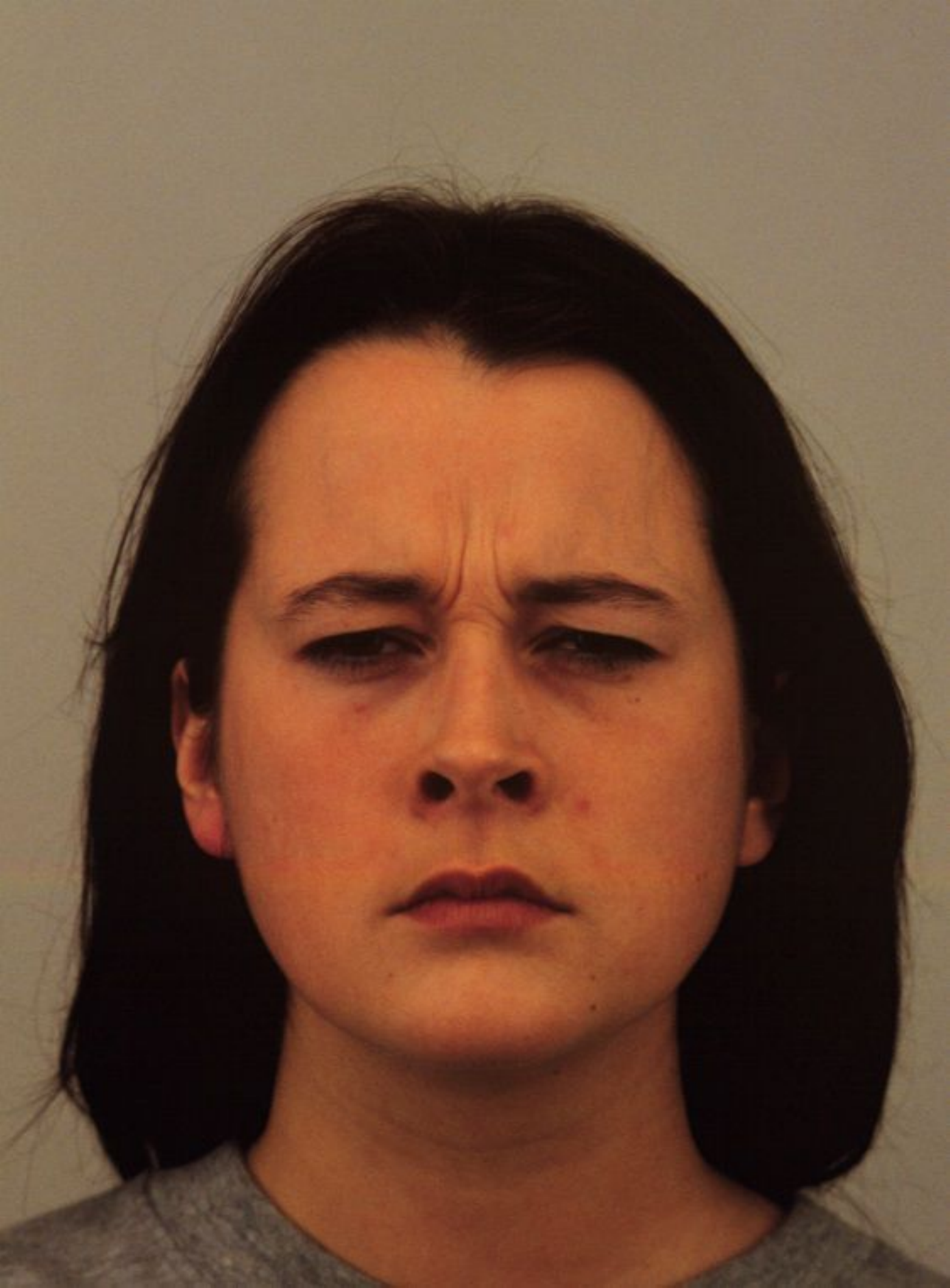}
\hfill
\includegraphics[width=0.158\linewidth]{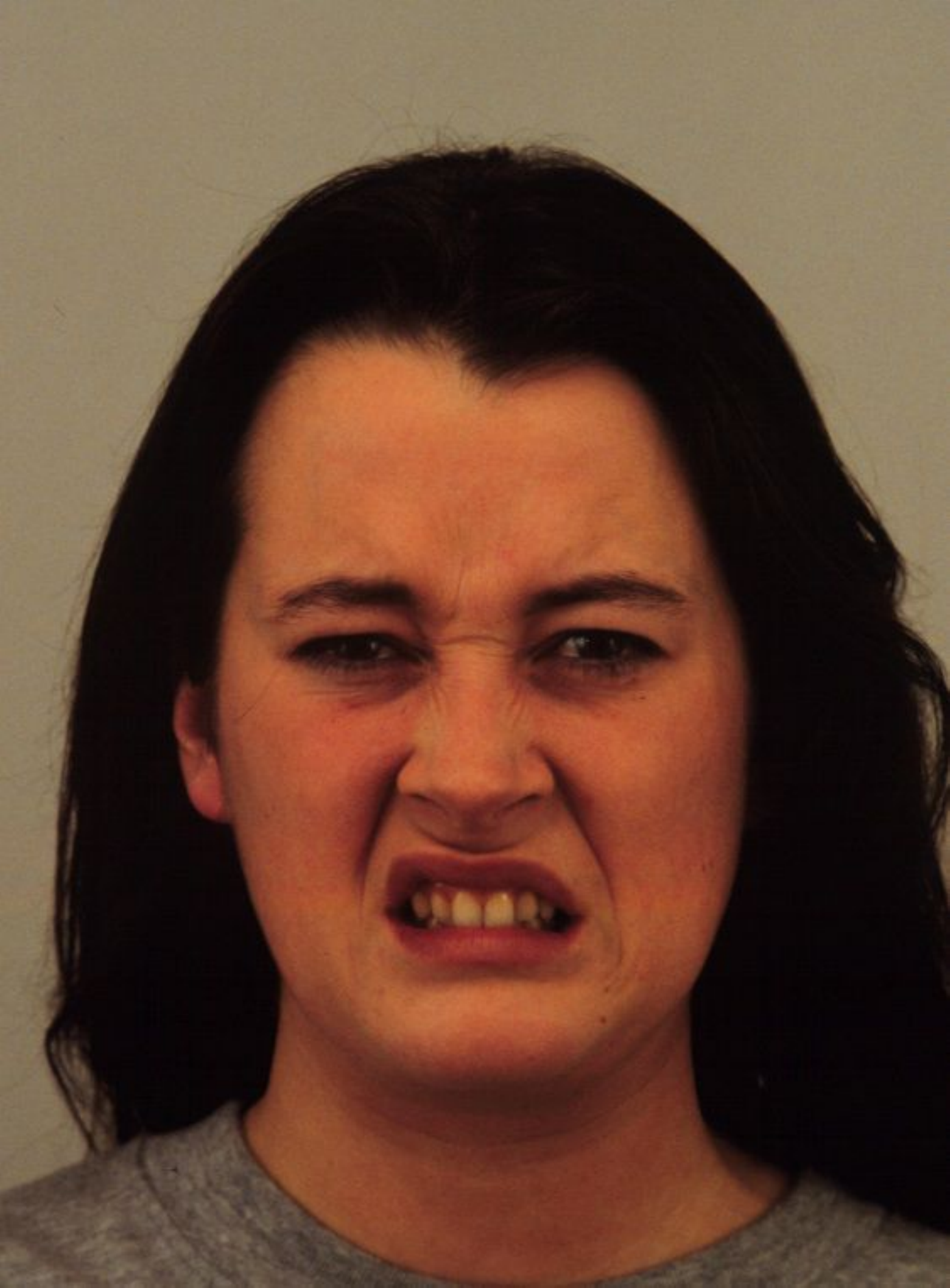}
\hfill
\includegraphics[width=0.158\linewidth]{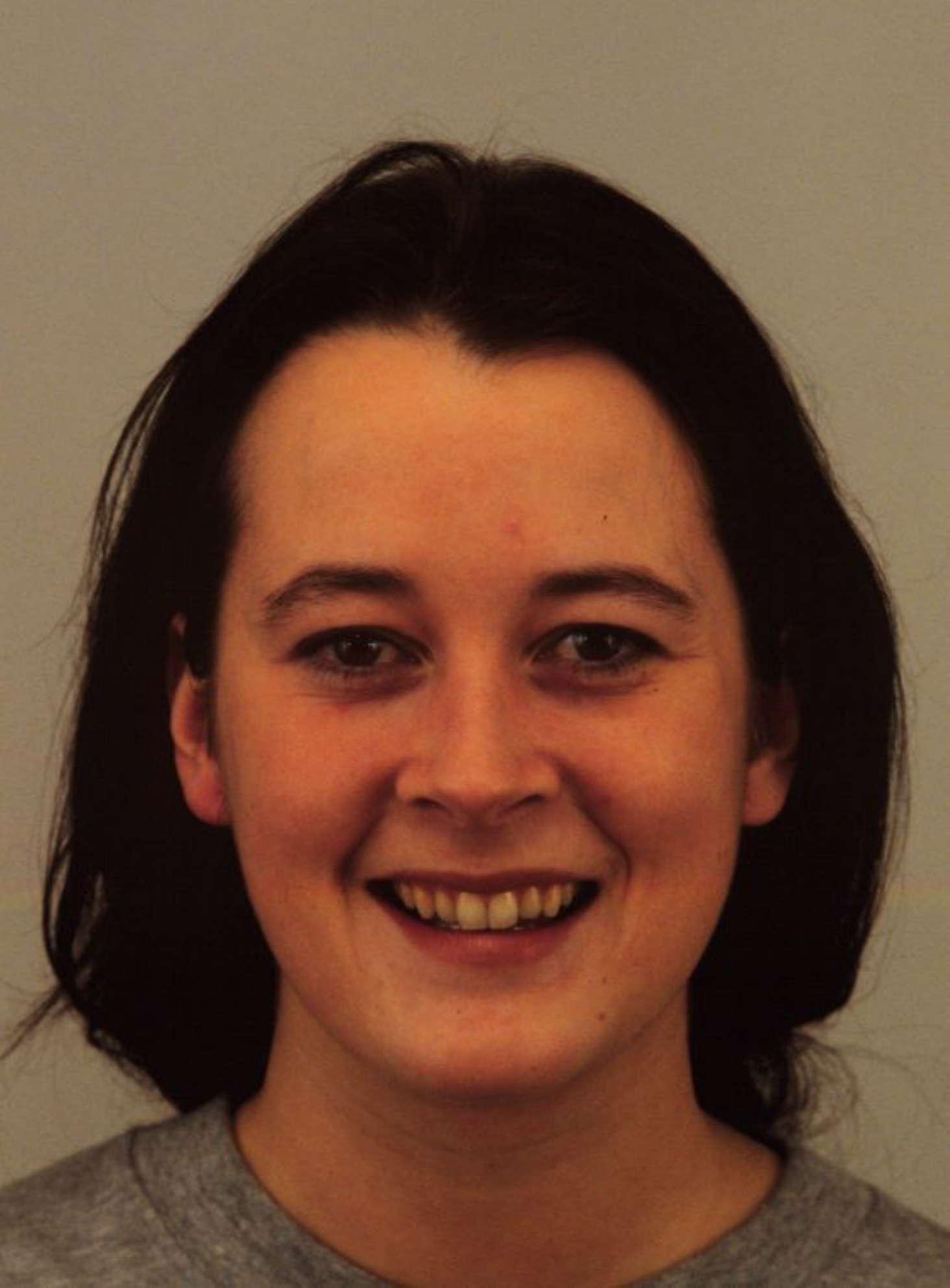}
\hfill
\includegraphics[width=0.158\linewidth]{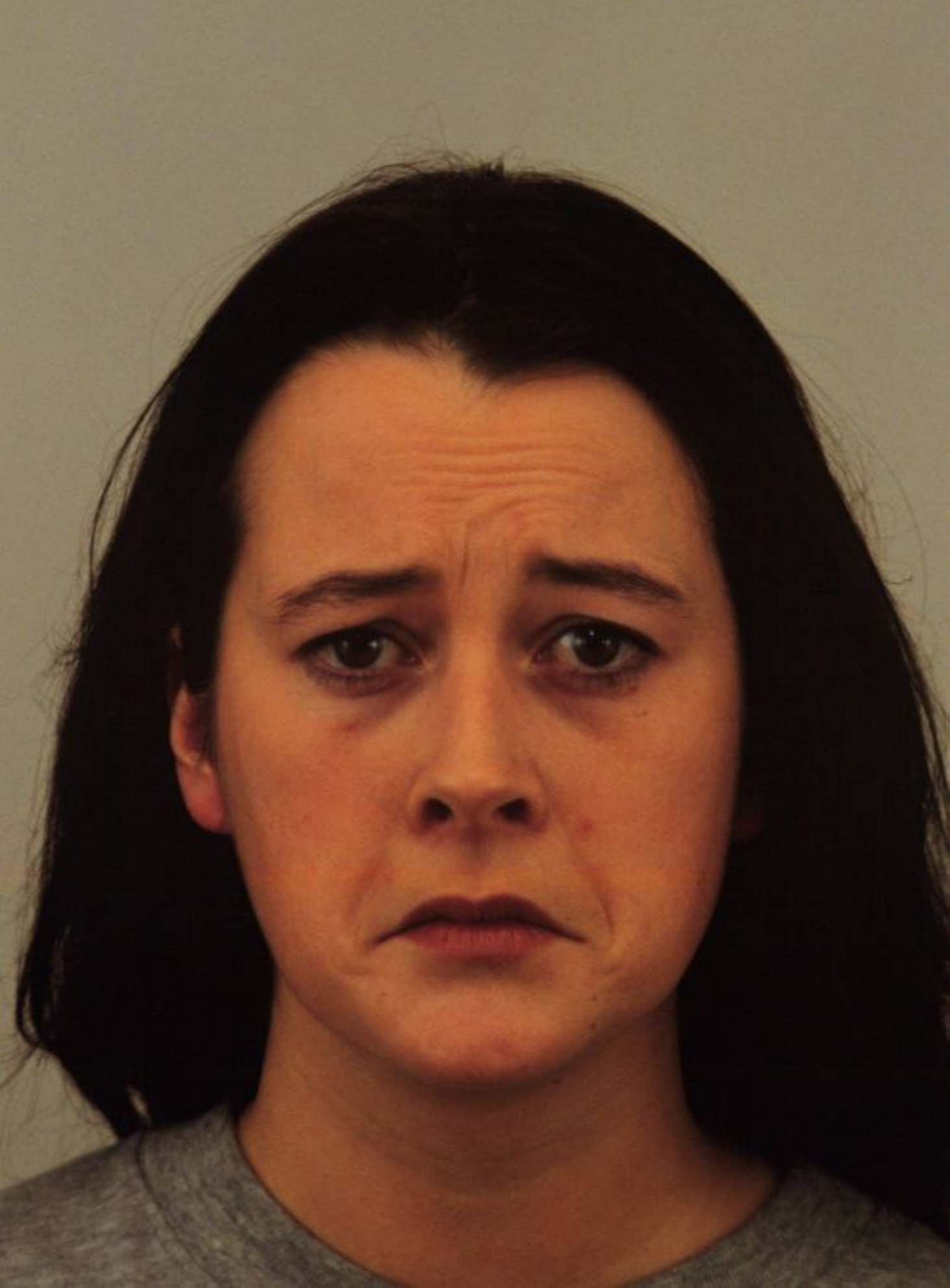}
\hfill
\includegraphics[width=0.158\linewidth]{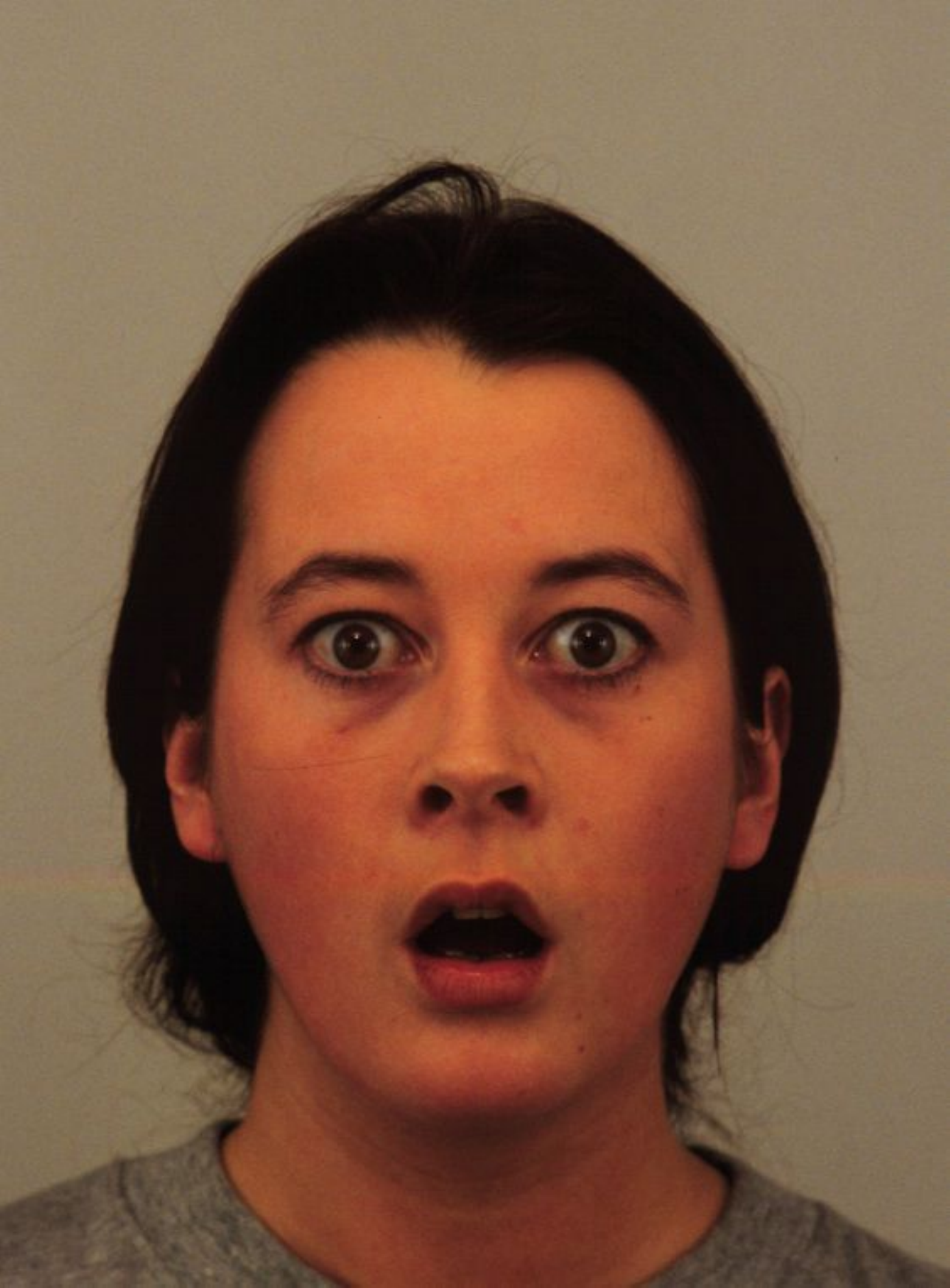}\\
\includegraphics[width=0.158\linewidth]{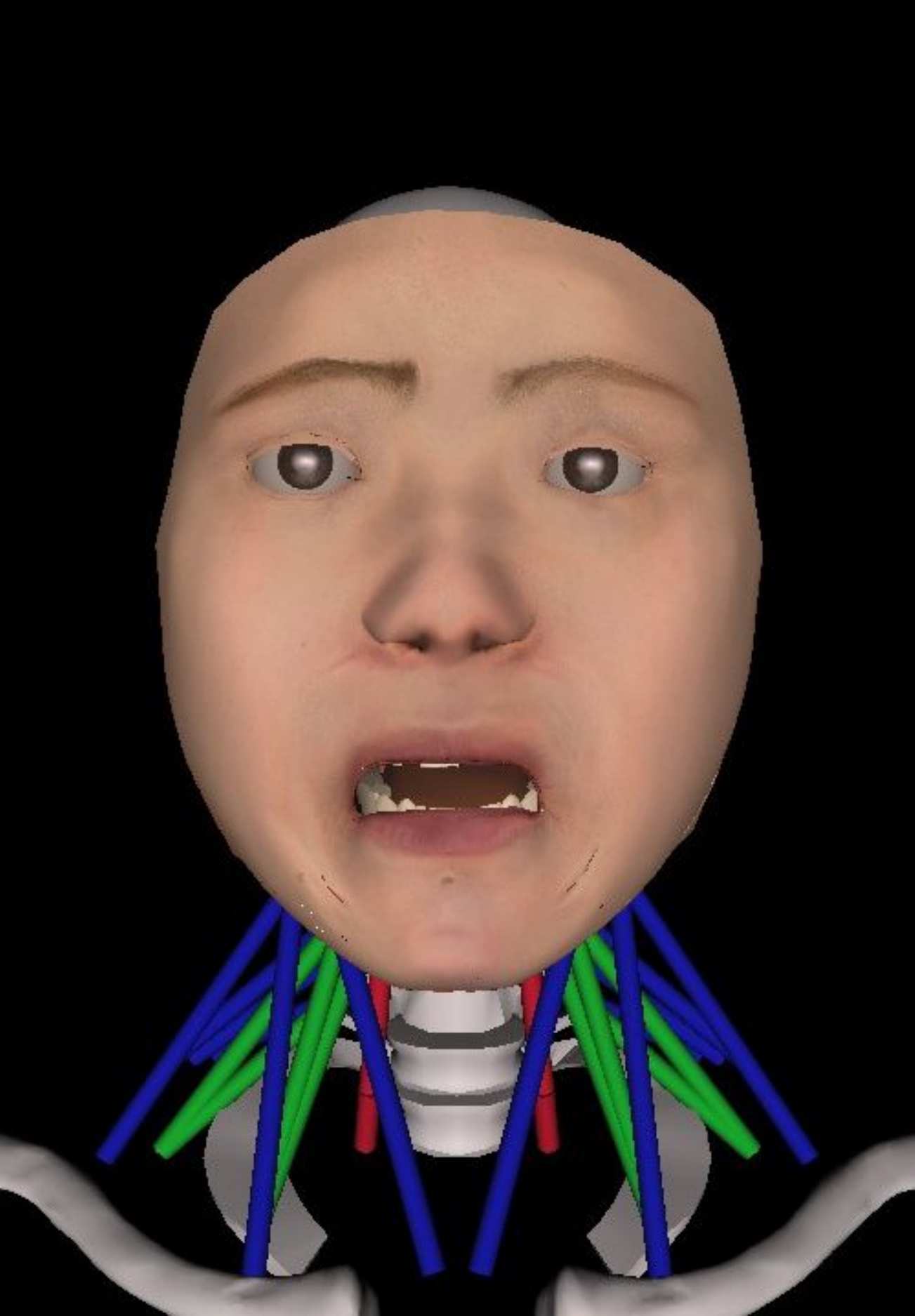}
\hfill
\includegraphics[width=0.158\linewidth]{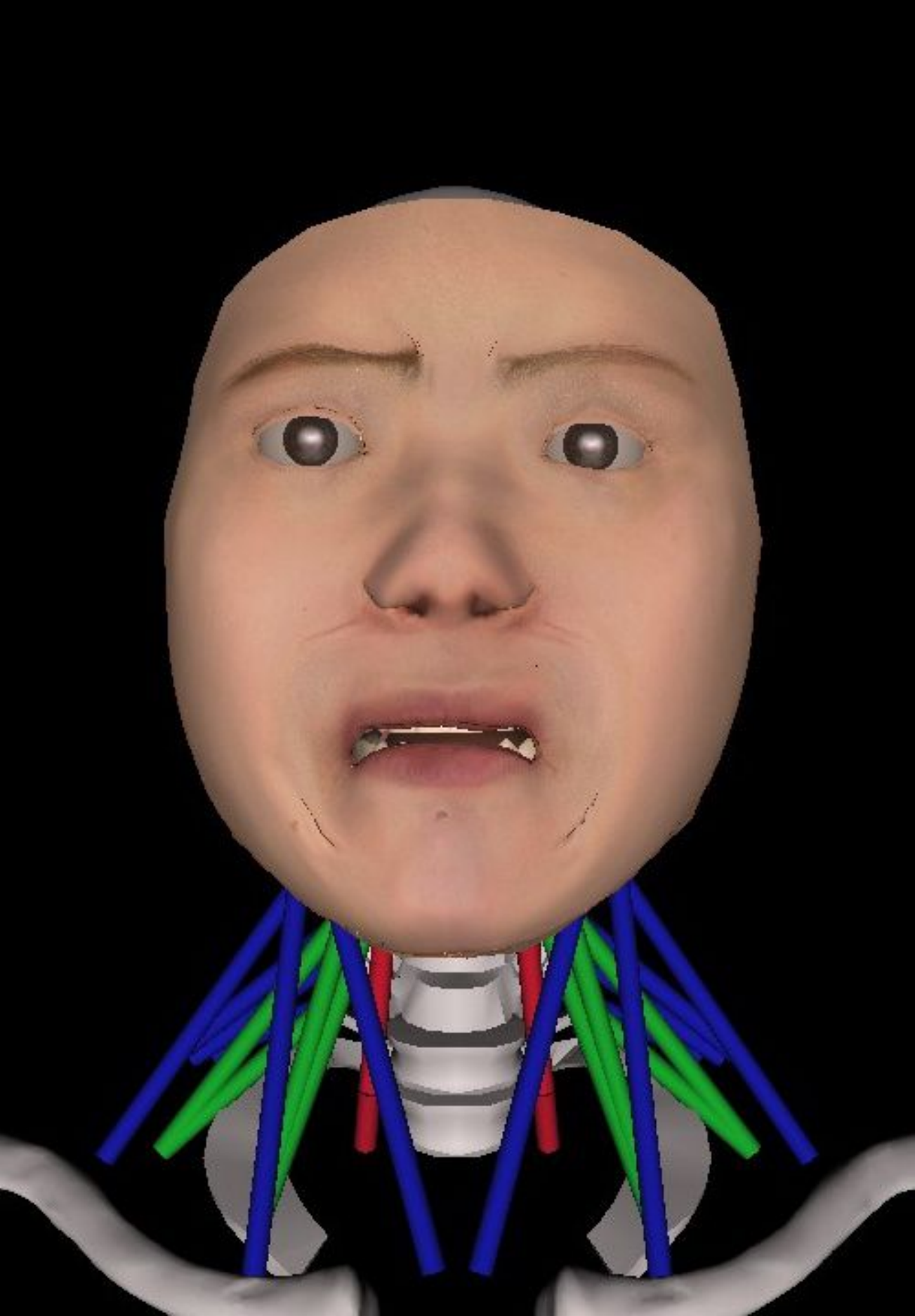}
\hfill
\includegraphics[width=0.158\linewidth]{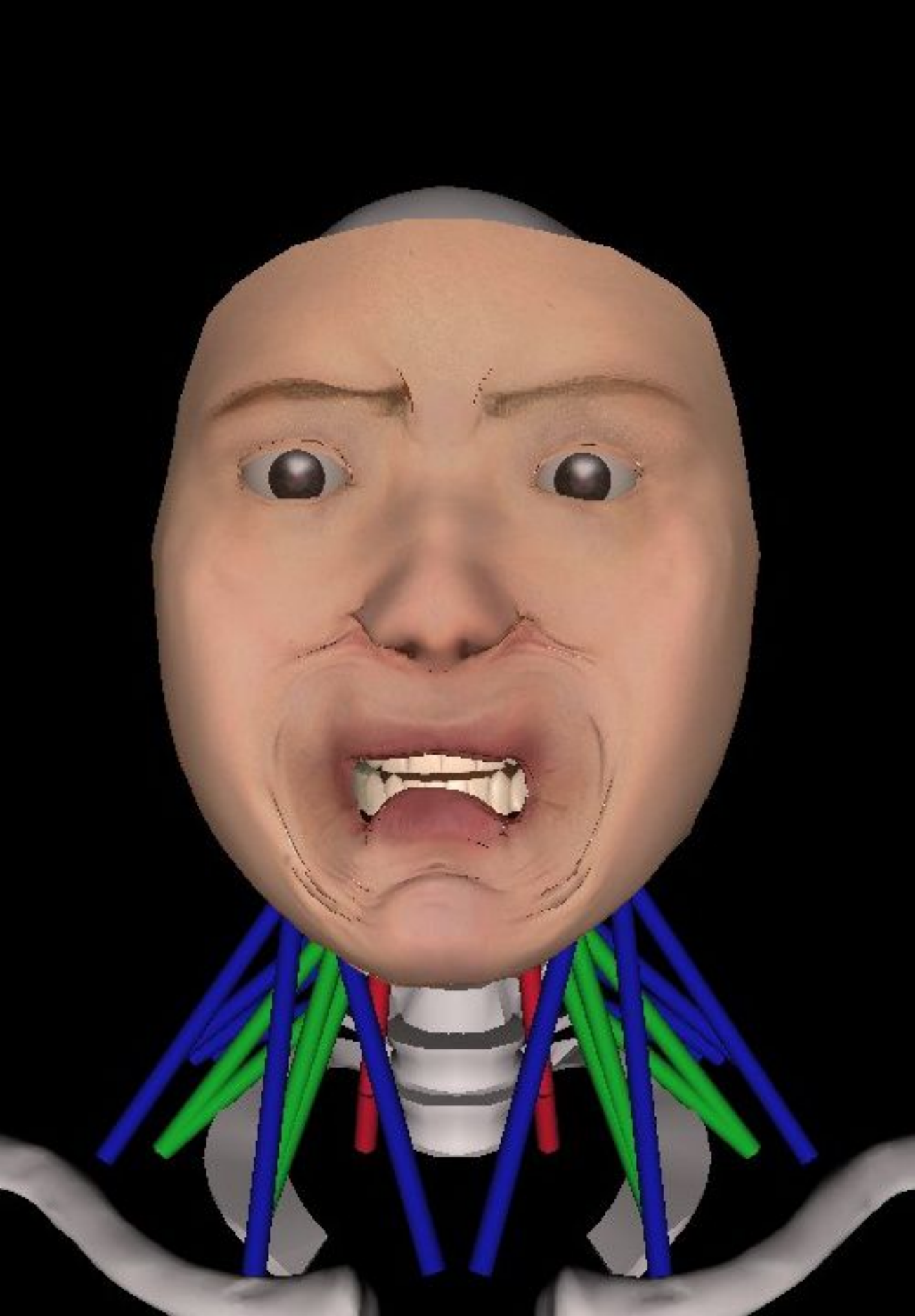}
\hfill
\includegraphics[width=0.158\linewidth]{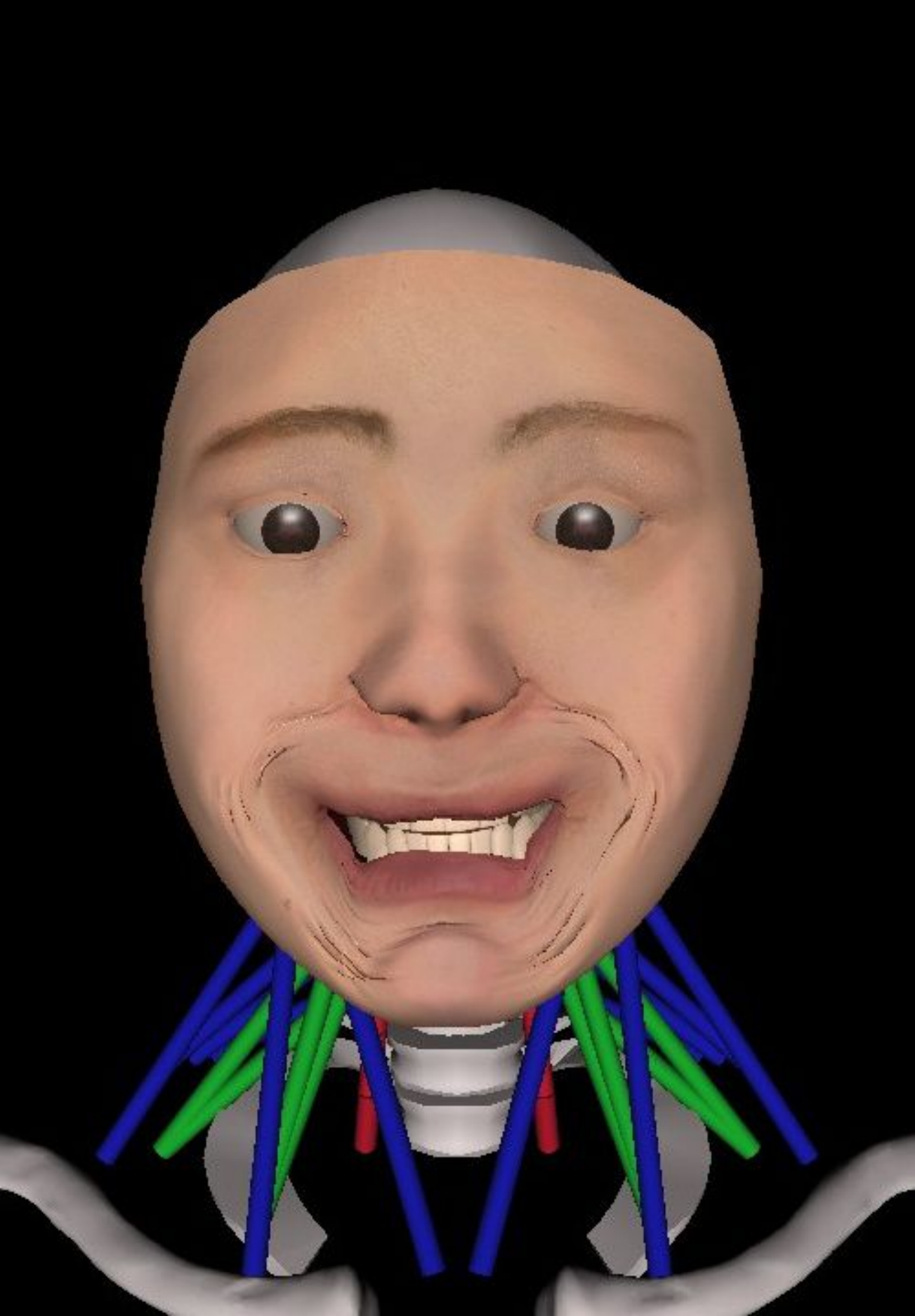}
\hfill
\includegraphics[width=0.158\linewidth]{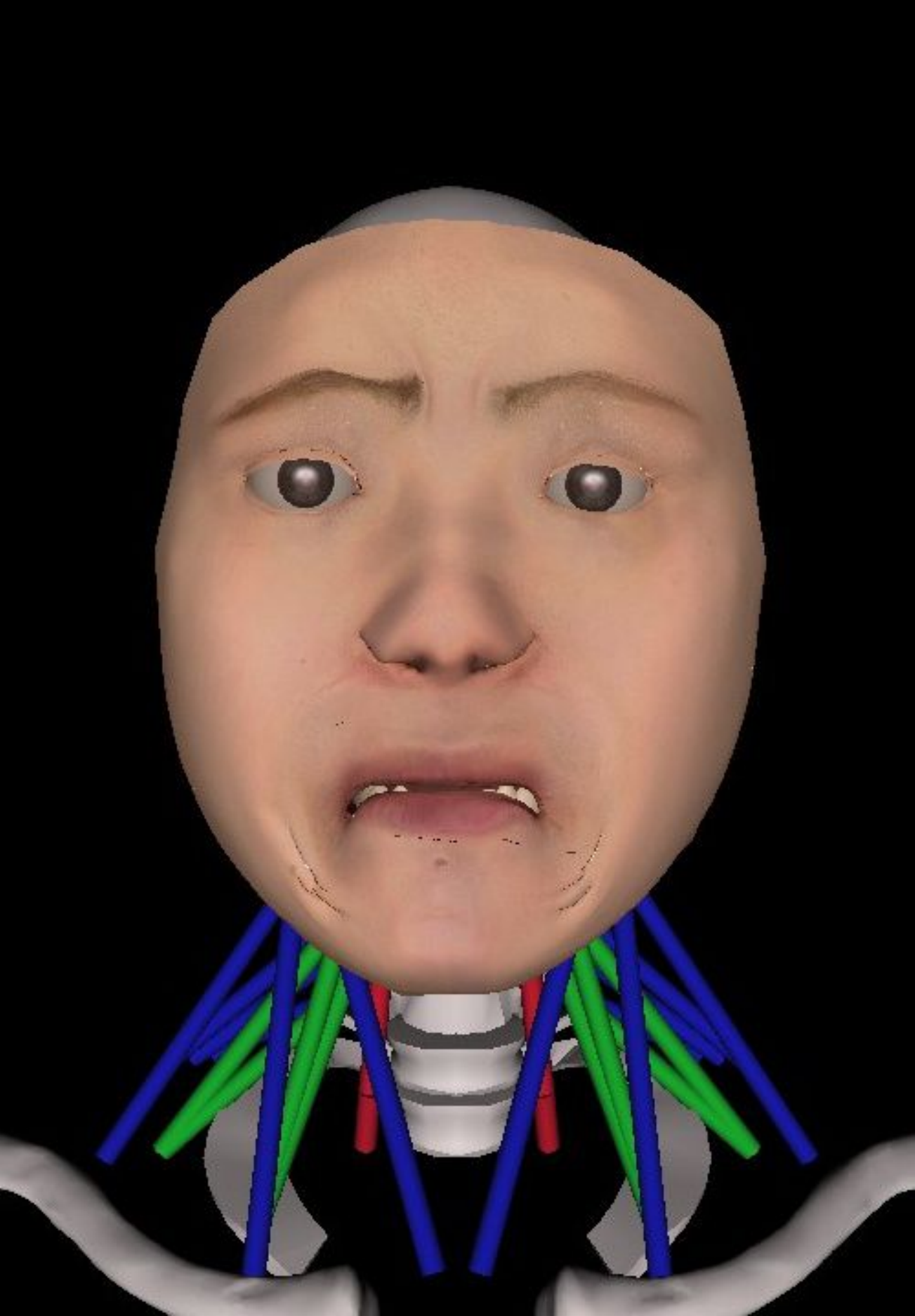}
\hfill
\includegraphics[width=0.158\linewidth]{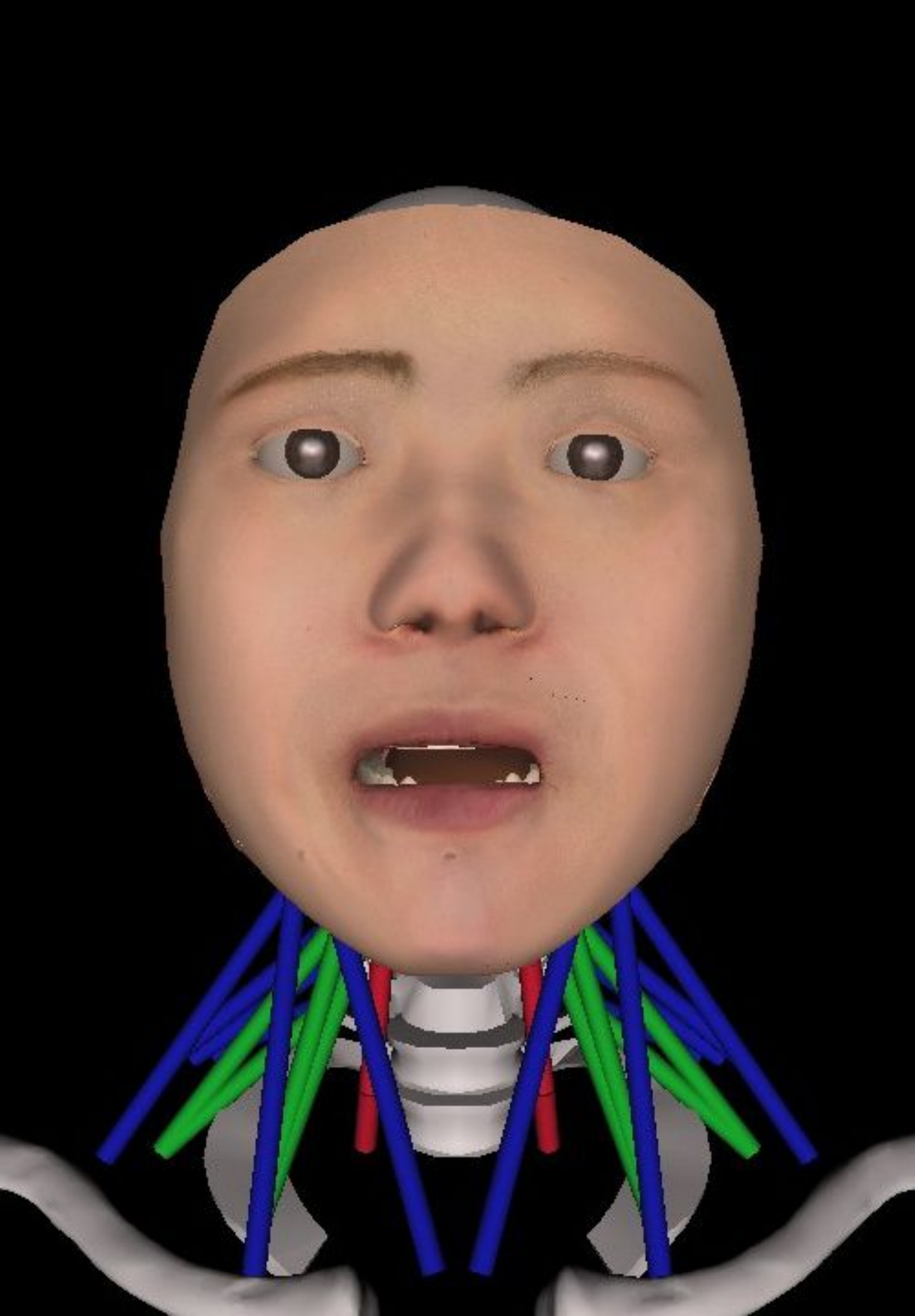}}\\[5pt]
\subcaptionbox{Subject 5}[\linewidth]{\includegraphics[width=0.158\linewidth]{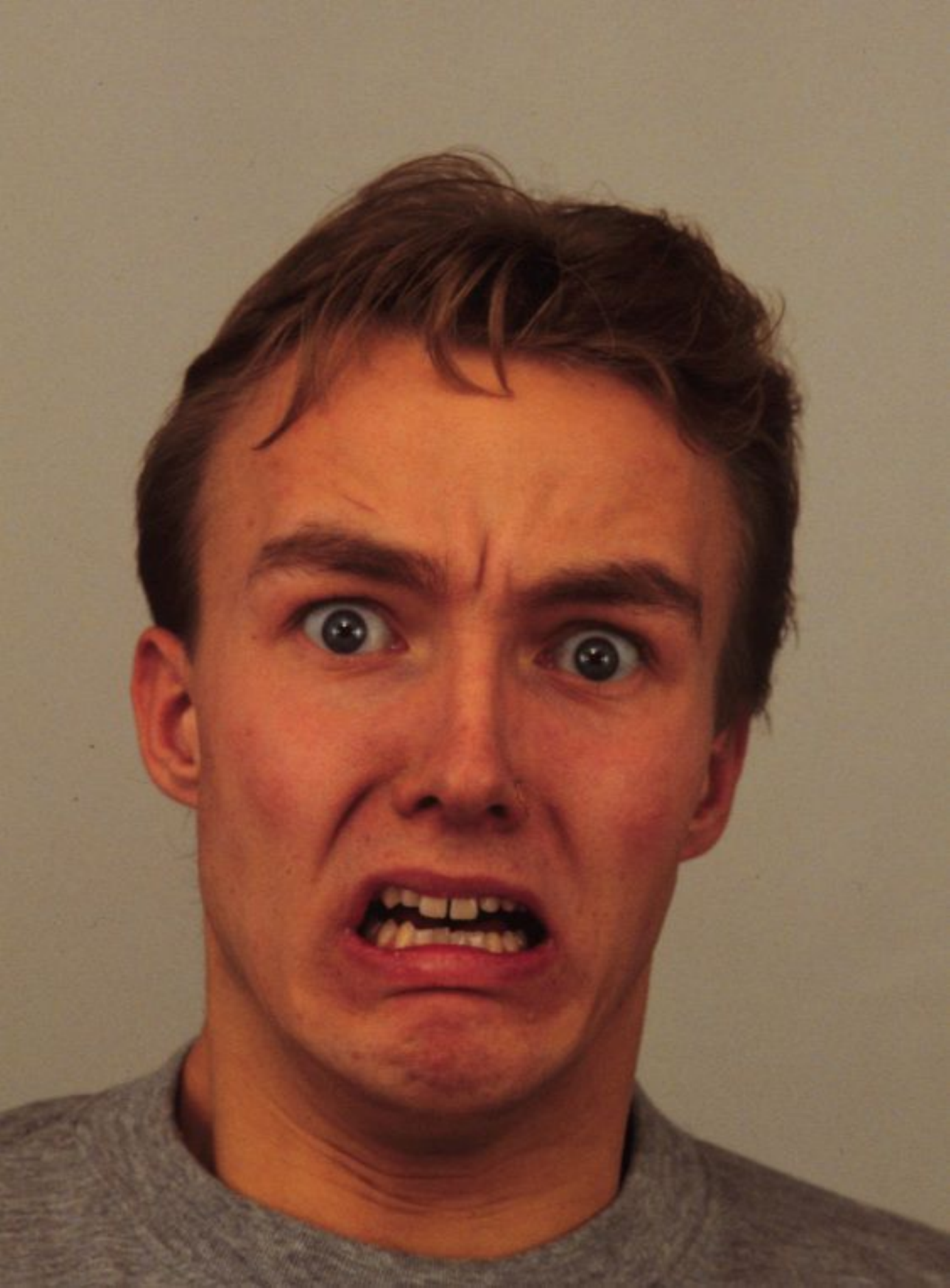}
\hfill
\includegraphics[width=0.158\linewidth]{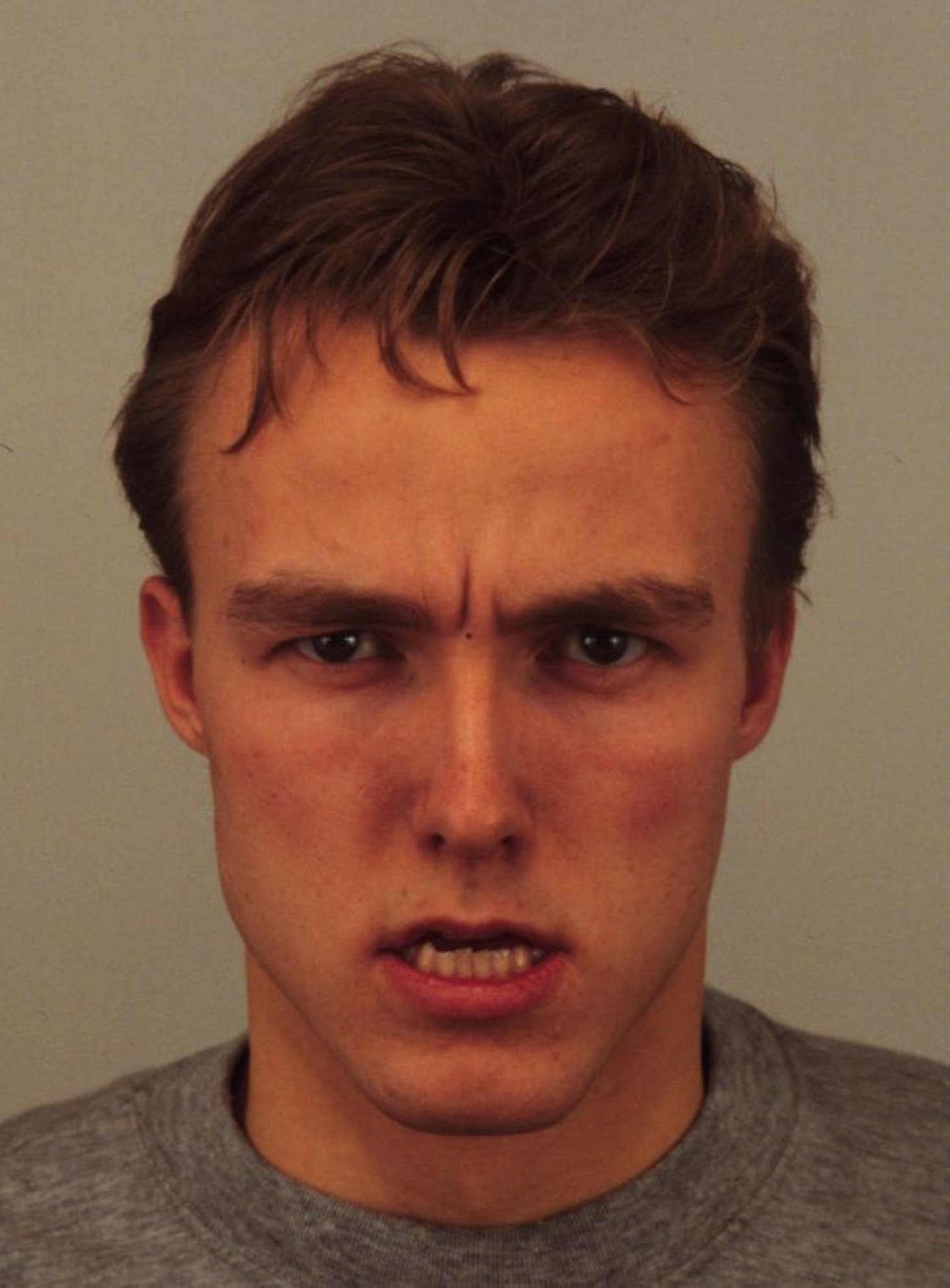}
\hfill
\includegraphics[width=0.158\linewidth]{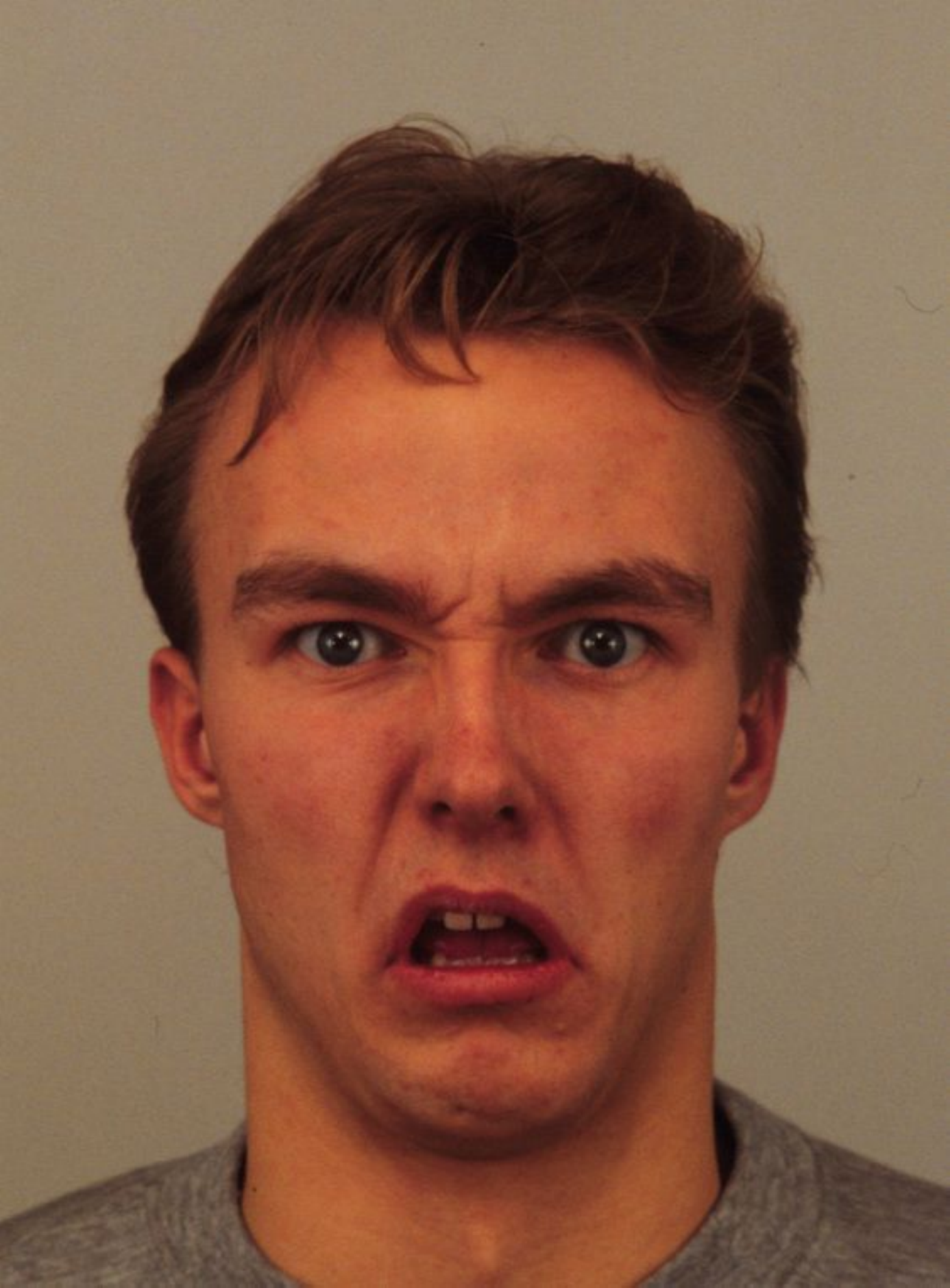}
\hfill
\includegraphics[width=0.158\linewidth]{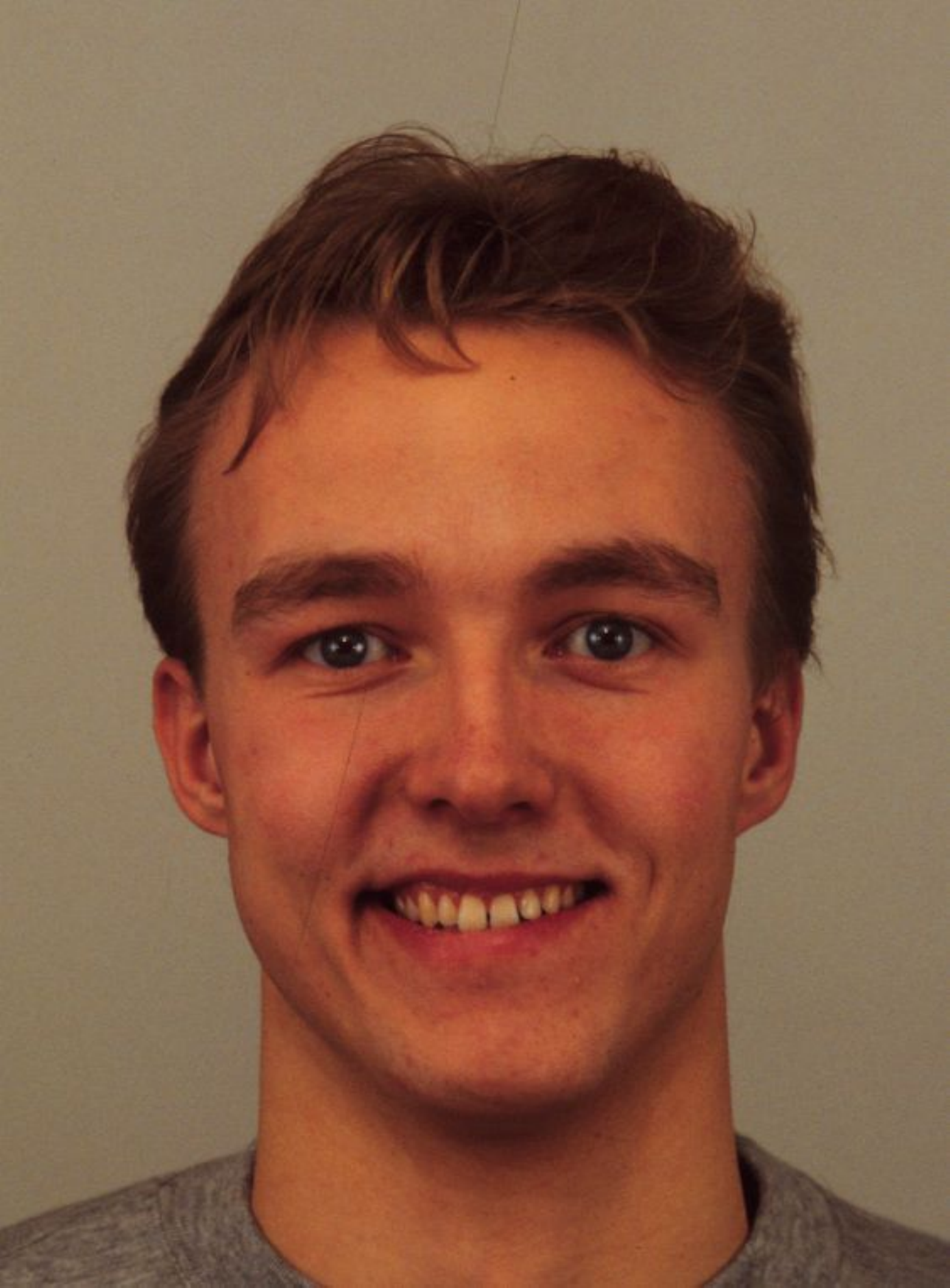}
\hfill
\includegraphics[width=0.158\linewidth]{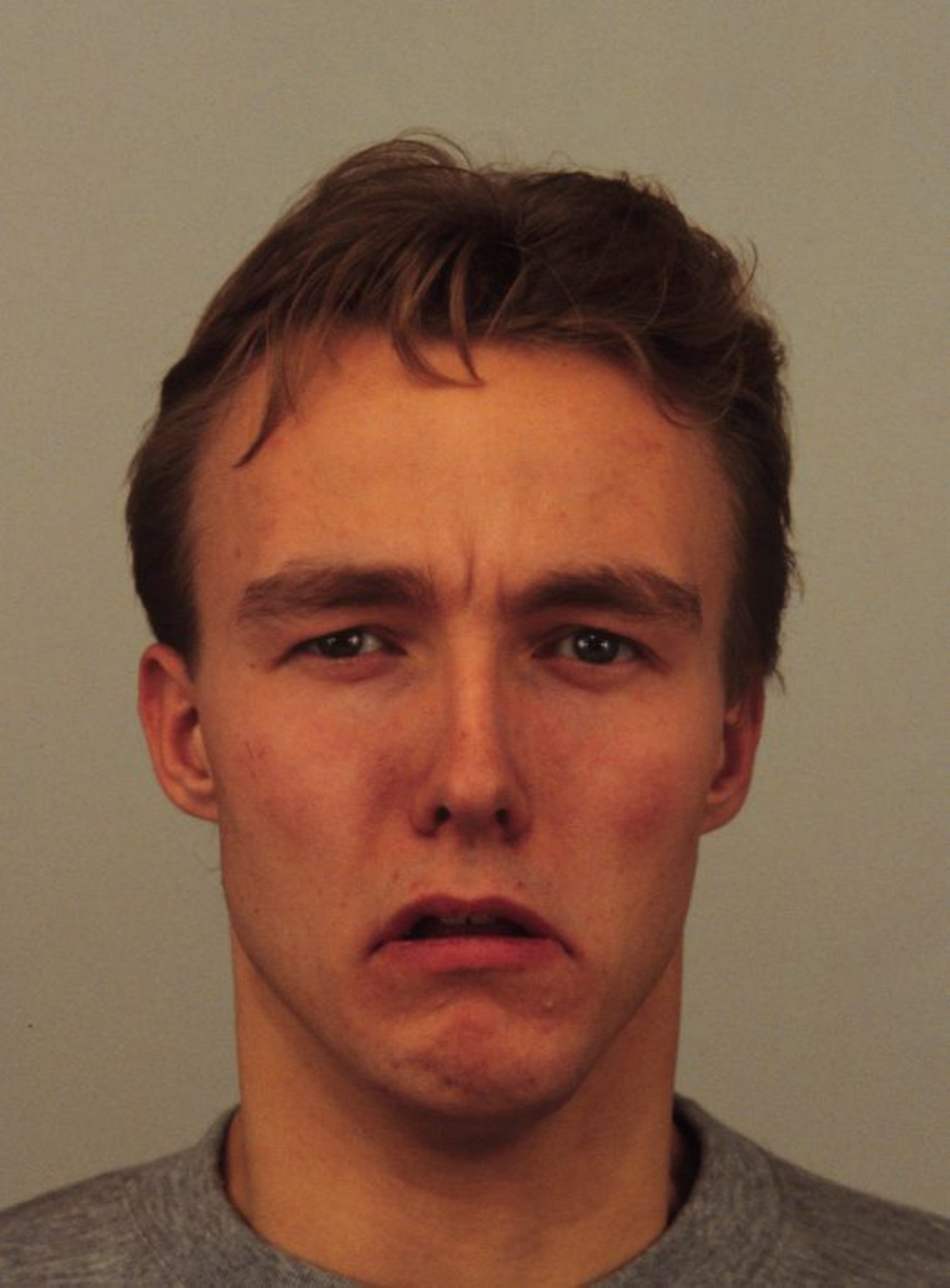}
\hfill
\includegraphics[width=0.158\linewidth]{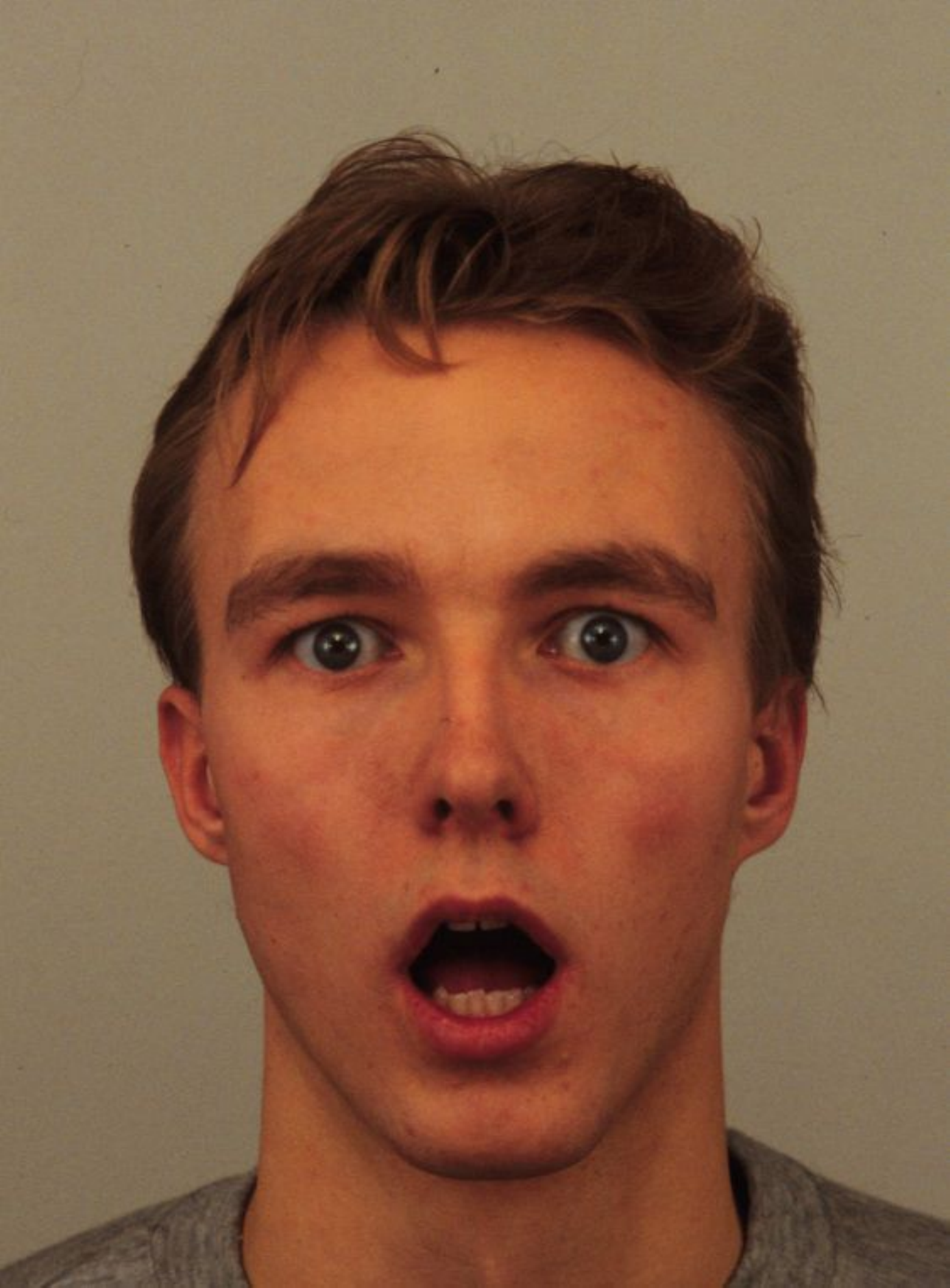}\\
\includegraphics[width=0.158\linewidth]{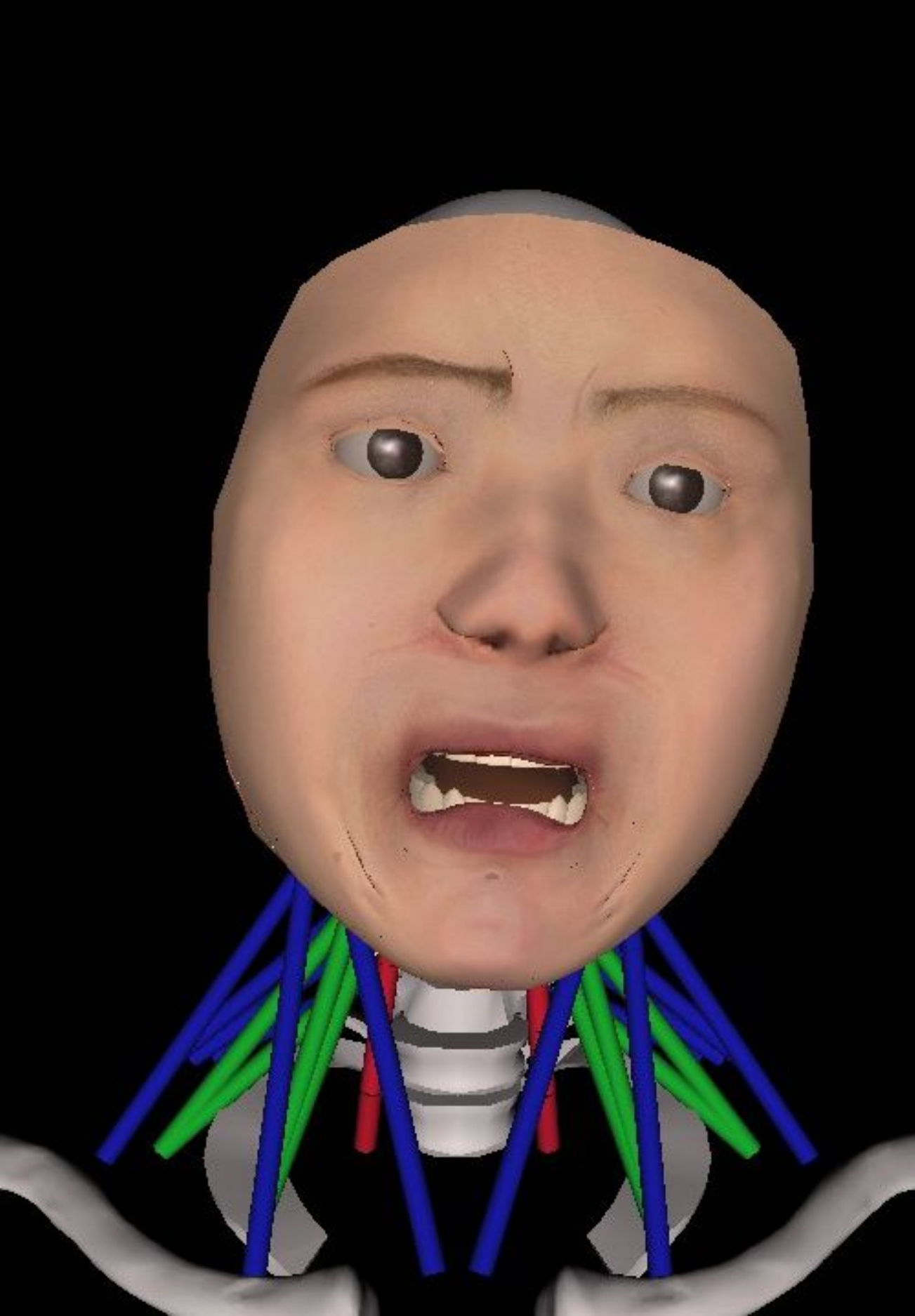}
\hfill
\includegraphics[width=0.158\linewidth]{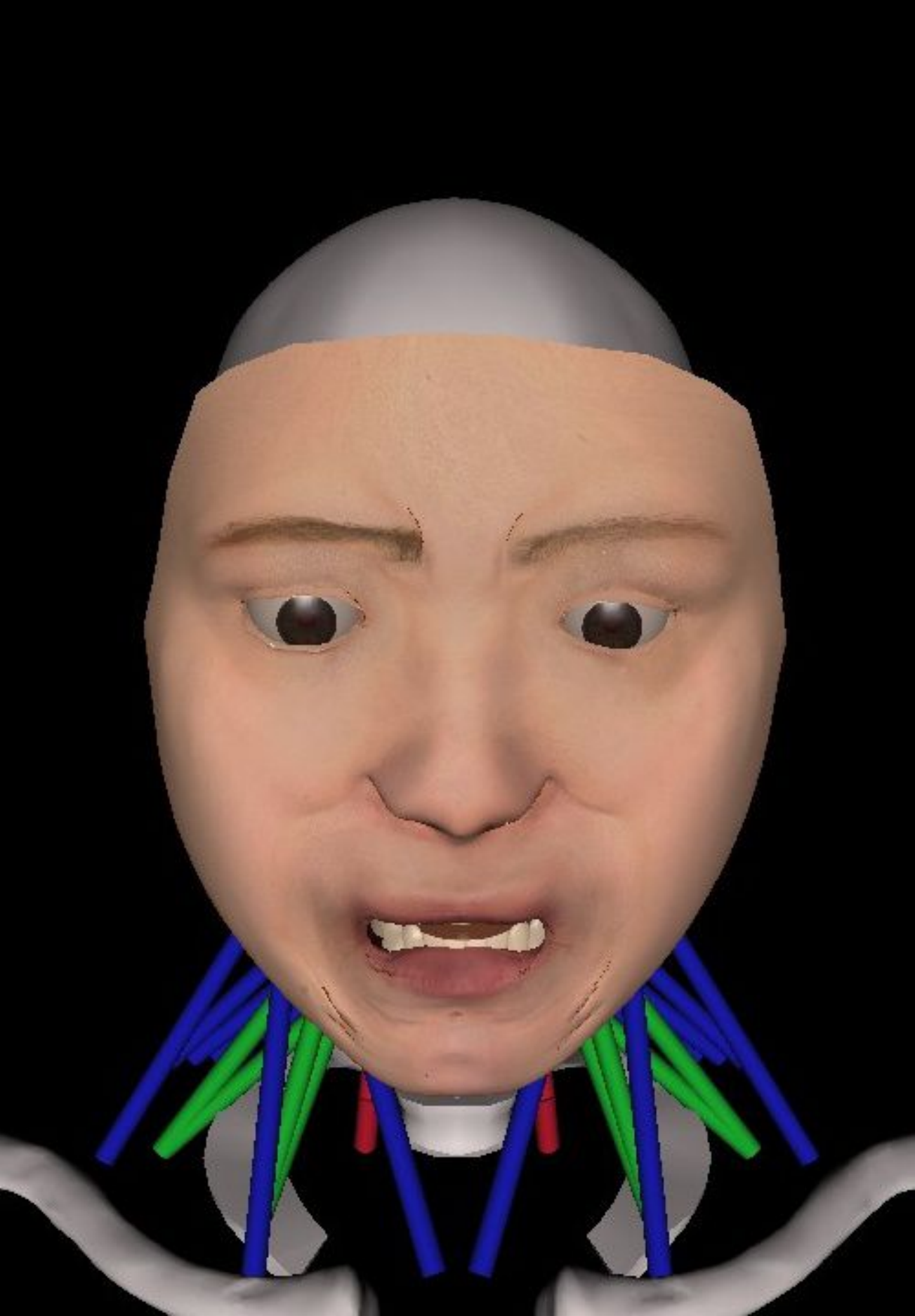}
\hfill
\includegraphics[width=0.158\linewidth]{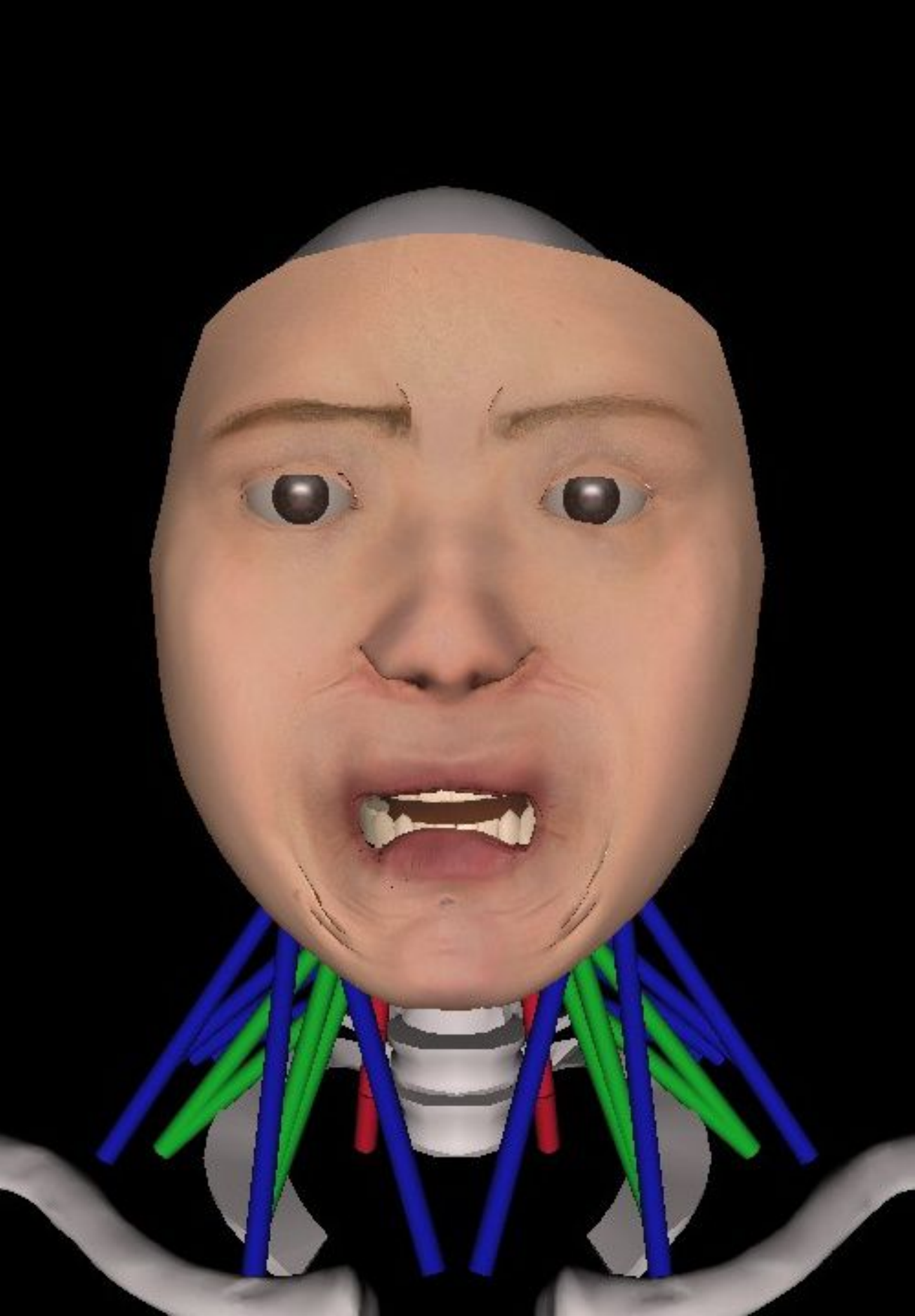}
\hfill
\includegraphics[width=0.158\linewidth]{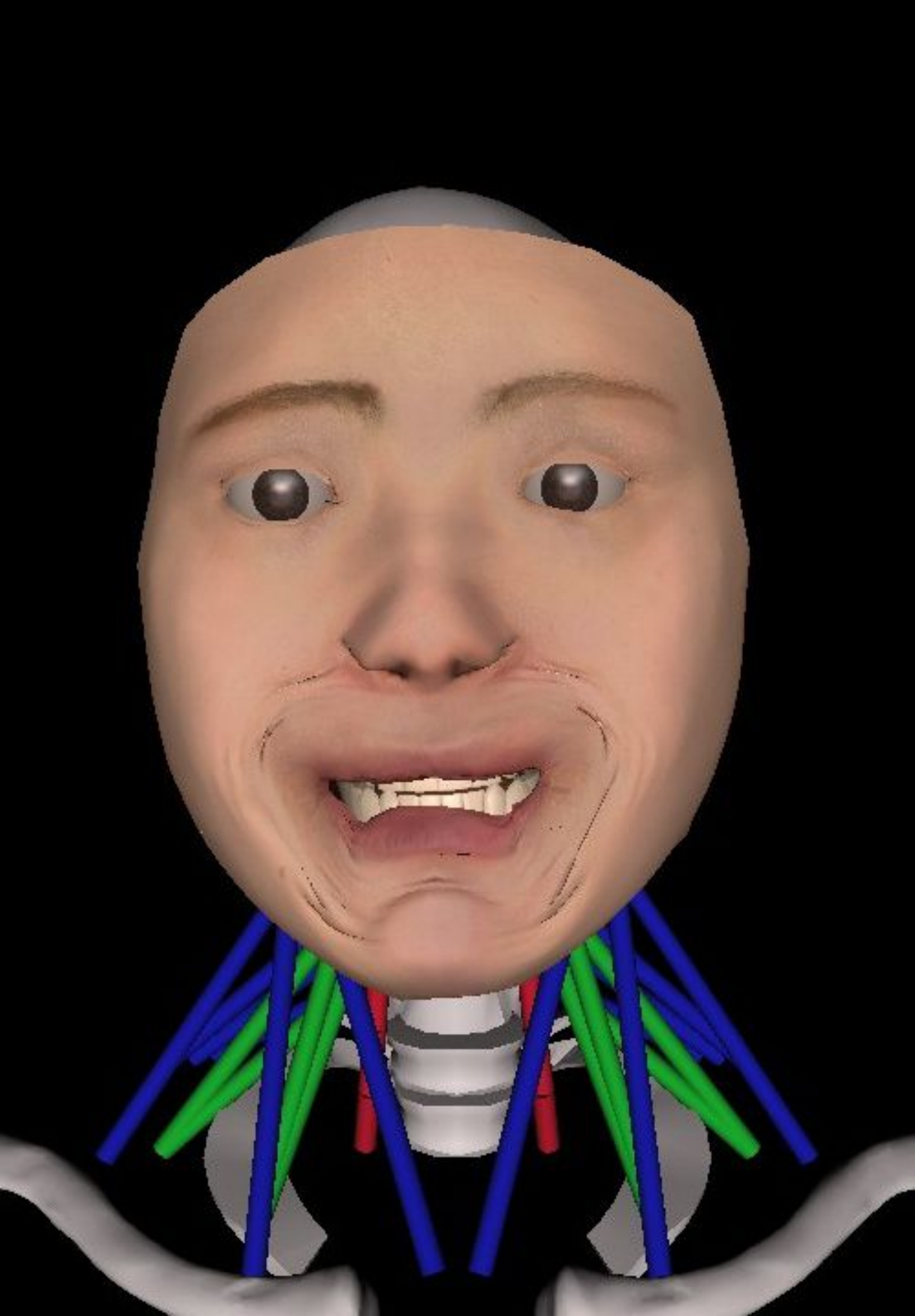}
\hfill
\includegraphics[width=0.158\linewidth]{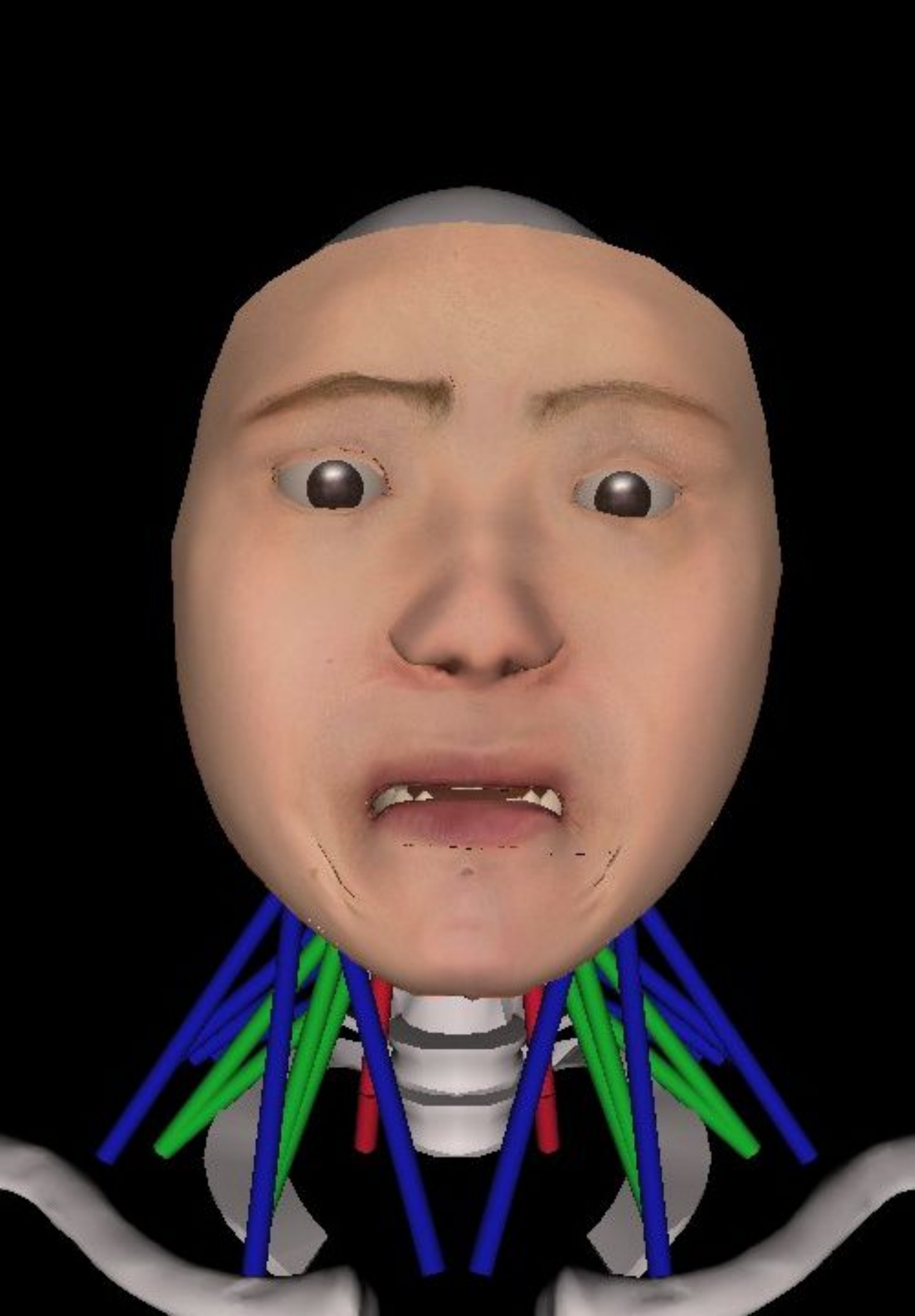}
\hfill
\includegraphics[width=0.158\linewidth]{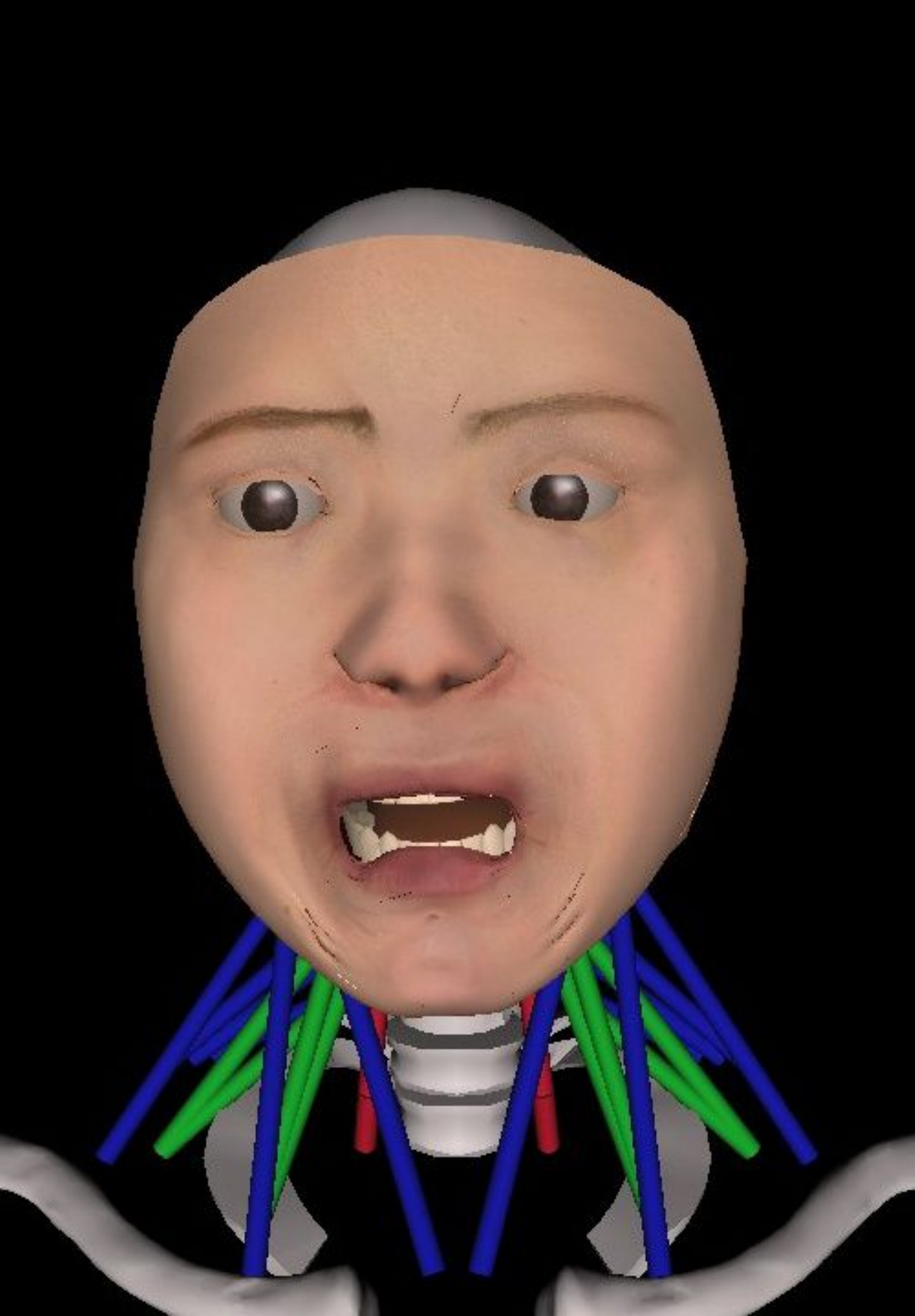}}
\caption{Transfer of facial expressions (fear, anger, disgust, joy,
sadness, surprise) and head poses from KDEF images to the
biomechanical face-head-neck model.}
\label{fig:recreations}
\end{figure}

\subsection{Facial Expression Transfer}

The results for the facial expression transfer for each of the
expressions is shown in Figure~\ref{fig:recreations}. We present
transfer results of a small sample of the KDEF dataset, selecting two
subjects (a female and a male) enacting all the basic expressions.

We also evaluate the transfer results by comparing the MSE of selected
AUs in Table~\ref{table:mse_compare}. We calculate the average MSE by
AUs over all transferred expressions and their corresponding reference
data, then compute the mean value of the average MSEs over all AUs in
each experimental setting. We report such mean values together with 8
randomly chosen AUs. The normalization step decreases the MSEs of most
selected AUs and yields results with higher AU similarity. 

\begin{figure} \centering
\includegraphics[width=0.16\linewidth]{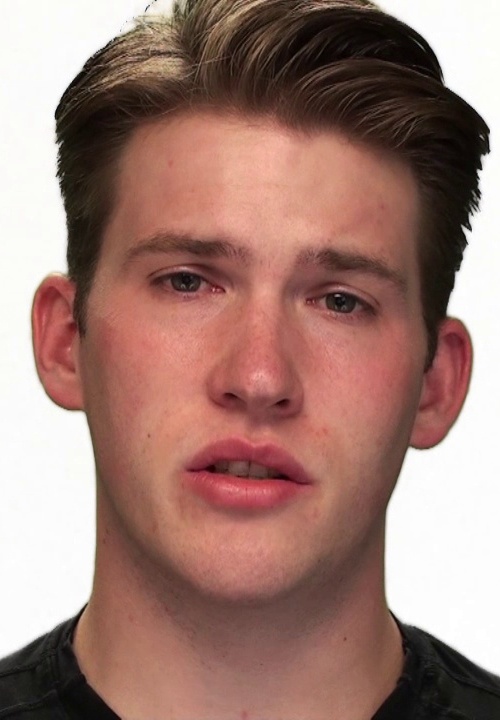}
\includegraphics[width=0.16\linewidth]{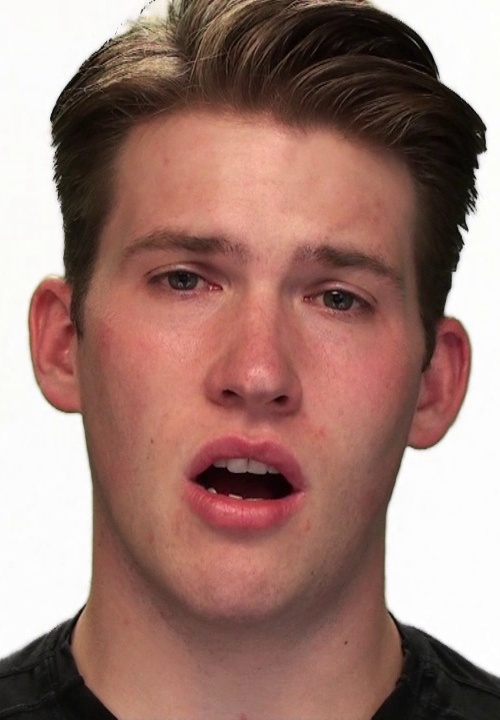}
\includegraphics[width=0.16\linewidth]{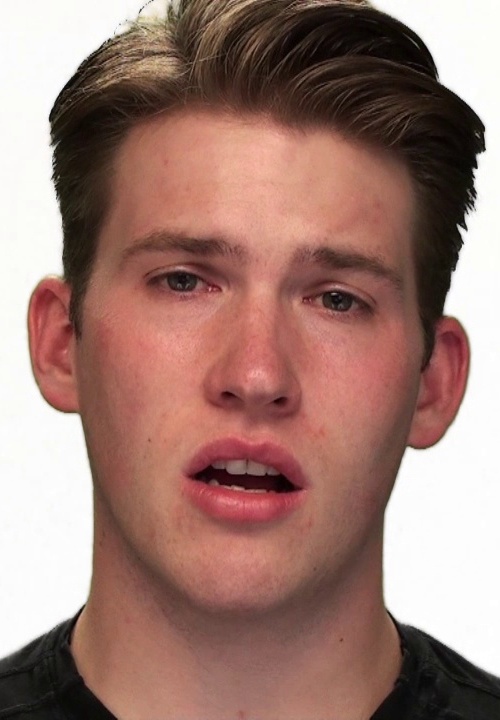}
\includegraphics[width=0.16\linewidth]{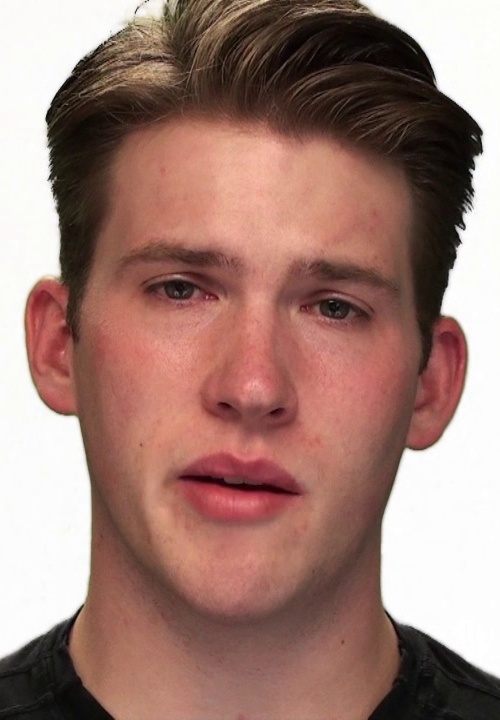}
\includegraphics[width=0.16\linewidth]{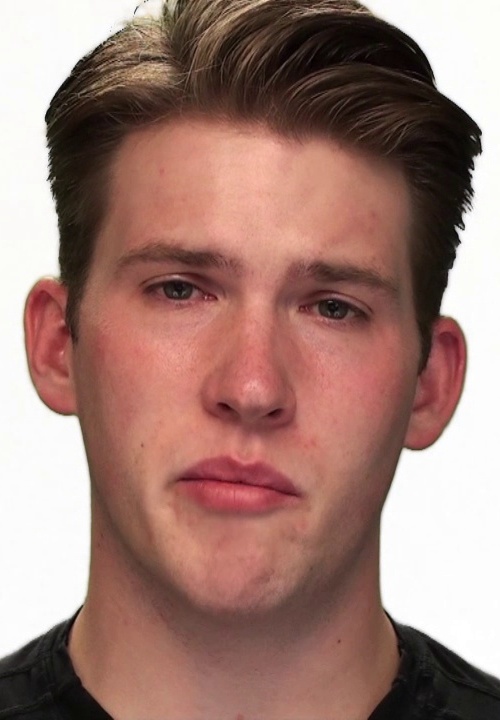}\\[2pt]
\includegraphics[width=0.16\linewidth]{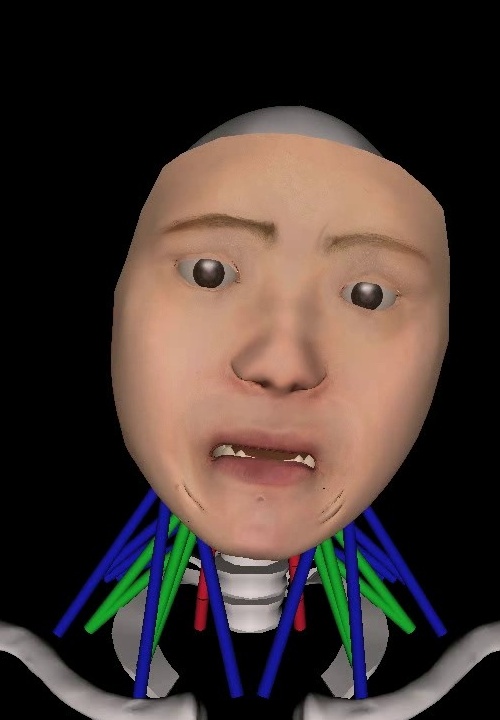}
\includegraphics[width=0.16\linewidth]{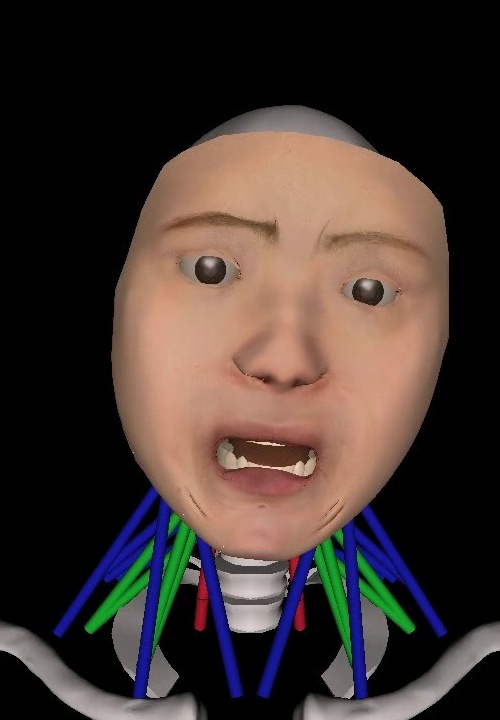}
\includegraphics[width=0.16\linewidth]{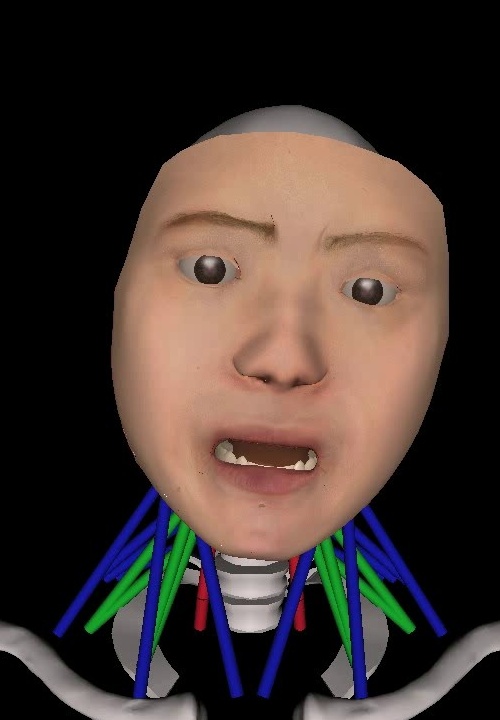}
\includegraphics[width=0.16\linewidth]{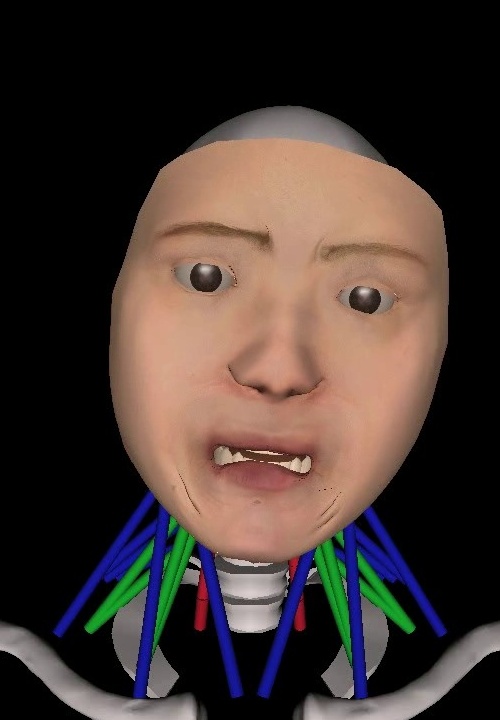}
\includegraphics[width=0.16\linewidth]{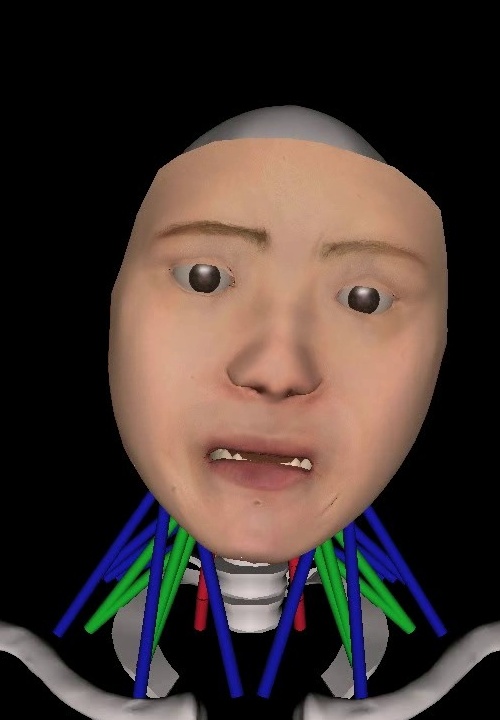}
\caption{Facial expression and head pose transfer from a video image
sequence from the RAVDESS dataset of a talking subject expressing
Sadness.}
\label{fig:RAVDESS_sequence_2}
\end{figure}

Figure~\ref{fig:RAVDESS_sequence_2} shows selected frames of the
transfer of expressions from a video in the RAVDESS dataset (best seen
in our supplemental video). The more subtle expressions performed by
subjects in the RAVDESS and CK videos result in lower average MSEs
compared with those for the KDEF image dataset.

\begin{table} \centering
\caption{Average MSE of selected AUs in the original data and transfer
results using different settings. Note that we randomly select 8 out
of 17 AUs output by OpenFace. The last column is the average MSE over
all the AUs in each setting.}
\label{table:mse_compare}
\resizebox{\linewidth}{!}{
\begin{tabular}{ l c c c c c c c c c} 
\toprule
Dataset & $\text{MSE}_{\text{AU1}}$ & $\text{MSE}_{\text{AU4}}$ & $\text{MSE}_{\text{AU6}}$ & $\text{MSE}_{\text{AU9}}$ & $\text{MSE}_{\text{AU12}}$ & $\text{MSE}_{\text{AU15}}$ & $\text{MSE}_{\text{AU20}}$ & $\text{MSE}_{\text{AU25}}$ & Average \\
\midrule
KDEF (unnormalized) & 0.191 &	0.088 &	0.186 &	0.133 &	0.011 &	0.155 &	0.129 &	0.092 &	0.157 \\ 
KDEF (normalized) & 0.118 &	0.064 &	0.186 &	0.068 &	0.028 &	0.109 &	0.146 &	0.052 &	0.129 \\
RAVDESS (video) & 0.012 &	0.222 &	0.107 &	0.044 &	0.083 &	0.016 &	0.077 &	0.036 &	0.086 \\
CK (video) & 0.038 &	0.054 &	0.059 &	0.015 &	0.008 &	0.023 &	0.057 &	0.034 &	0.048 \\
\bottomrule
\end{tabular}
}
\bigskip
\end{table}

\section{Conclusions and Future Work}

Our expression transfer approach is uniquely advantageous. First, it
is anatomically consistent as the biomechanical model emulates the
human cervicocephalic musculoskeletal system. Further, our approach is
based on the Facial Action Coding System, which is a widely adopted
representation of facial expressions, including in the computer
animation field. Finally, our approach is based on deep learning,
which can capture the complex, non-linear relation between the Action
Units of the FACS and associated facial muscle activations.

Additionally, our neural-network-based approach does not require any
manual data collection as the training data is generated directly
using the biomechanical model itself. This only needs to take place
once, in advance. Once trained offline using the synthesized training
data, the neural network can quickly transfer a large number of facial
expressions. Unlike popular face tracking software such as Faceware,
which requires manual calibration steps to transfer expressions of
different subjects, our proposed approach needs no additional
parameter adjustments to perform the same task.

Although we have achieved satisfactory transfer results for all the
basic facial expressions, note that the simulation and transfer of
complex mouth and eye movements lies outside the scope of our current
work. To improve the fidelity of the results, we plan to extend our
musculoskeletal model (e.g., by adding more muscles to help activate
facial AUs in a more anatomically accurate manner) and its associated
control system so as to explicitly control the lips and eyes, as well
as to transfer mixtures of expressions, subtle expressions, and
microexpressions, which the human face is capable of producing.

\begin{acks}
We thank CyberAgent, Inc., for providing the high resolution texture
for the facial mesh that we used in our model. We also thank the
curators of RAVDESS, CK/CK+, and KDEF for granting us permission to
use their datasets.
\end{acks}

\appendix

\section{Additional Details of the Biomechanical Model}
\label{app:details}

\subsection{Skeleton}

The skeletal structure of our model contains 9 links and is modeled as
an articulated rigid-body system. The bone links are built on the base
link, which is set to be immobile. Among the 8 movable bones, there
are seven cervical bones, named C1 to C7, and the skull. The human
neck skeleton system contains soft tissue between vertebrae and the
cervical spine, which enables motion with 6 degrees of freedom (3 for
translation and 3 for rotation) for the bones. We simplify each bone
joint to have just 3 rotational degrees of freedom.

The equation of motion of the rigid-body system can be written as
\begin{equation}
\vec{M}(\vec{q}) \ddot{\vec{q}} + \vec{C}(\vec{q}, \dot{\vec{q}}) + 
\vec{K}_s \vec{q} + \vec{K}_d \dot{\vec{q}} =
\vec{P}(\vec{q}) \vec{f}_m + \vec{J}(\vec{q})\tran \vec{f}_\text{ext},
\end{equation}
where $\vec{q}$, $\dot{\vec{q}}$, and $\ddot{\vec{q}}$ are
24-dimensional vectors representing the joint angles, the angular
velocities, and the angular accelerations, respectively. $\vec{M}$ is
the mass matrix, and $\vec{C}$ represents the internal forces among
the system, including Coriolis forces, gravity, and forces from
connecting tissues. The moment arm matrix $\vec{P}$ maps the muscle
force $\vec{f}_m$ (contractile muscle force $\vec{f}_C$ and passive
muscle force $\vec{f}_P$) to the related joint torques, while the
Jacobian matrix $\vec{J}$ transforms the applied external force
$\vec{f}_\text{ext}$ to torques. The $\vec{K}_s \vec{q} + \vec{K}_d
\dot{\vec{q}}$ term represents the rotational damping springs that we
attach to the joints in order to simulate the stiffness of the
inter-vertebral discs. We can alternatively write the torque from the
spring as:
\begin{equation}
\tau_s = -k_s(q-q_0)-k_d\dot{q},
\end{equation}
where $q$ is the current joint angle in the generalized coordinates,
and $q_0$ is the joint angle in the natural pose (resting angle). The
$k_s$ and $k_d$ are the stiffness and damping coefficients of the
spring, respectively.
 
\subsection{Muscles}

We use a modified version of the Hill-type muscle model
\cite{ng2001anatomically}, which is a good balance of biomechanical
accuracy and computational efficiency. The muscle force
$f_m={f}_P+{f}_C$ has two sources. The passive element ${f}_P$
generates a restoring force due to the muscle elasticity, which
constrains the muscle deformation passively. The passive muscle force
is represented as
\begin{equation}
f_P=\max(0, k_s(\exp(k_c e) -1) + k_d \dot{e}),
\end{equation}
where $k_s$ and $k_d$ are the stiffness and damping coefficients of
the above uni-axial exponential spring model. $e$ is the strain of the
muscle and $\dot{e}$ is the strain rate. We can calculate them using
$e=(l-l_0)/l_0$ and $\dot{e}=\dot{l}/l_0$, respectively, where $l$ and
$l_0$ are the muscle length and muscle resting length.

\begin{figure} \centering
\subcaptionbox{Force-length relationship \label{fig:1a}}
{\includegraphics[width=0.4\linewidth]{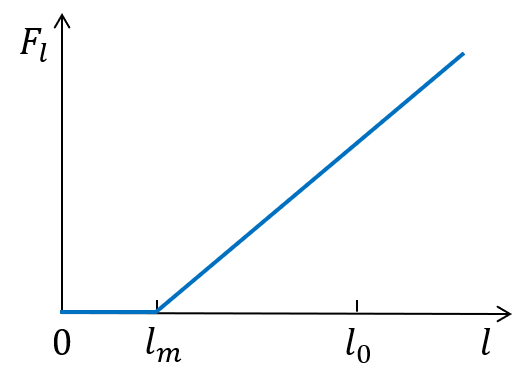}}
\qquad
\subcaptionbox{Force-velocity relationship \label{fig:1b}}
{\includegraphics[width=0.4\linewidth]{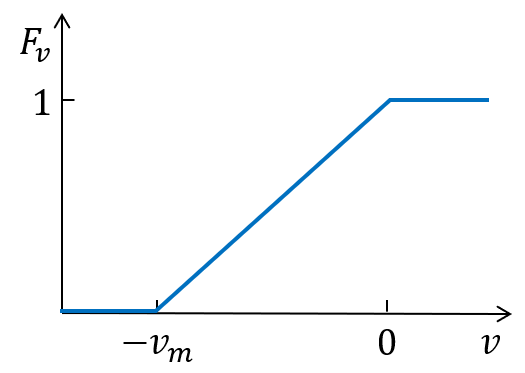}}
\caption{The force relationships of Hill type muscle
model.}
\label{fig:Fl_and_Fv}
\end{figure}

The contractile element $f_C$ generates the proactive contractile
force of the muscle, which is proportionate to the \emph{activation
level} of the muscle:
\begin{equation}
\begin{split}
f_C &= a F_l(l) F_v(\dot{l}),\\
F_l(l) &= \max(0, k_{max}(l-l_m)), \\
F_v(\dot{l}) &= \max(0, 1+{\min(\dot{l}, 0)}/{v_m}),
\end{split}
\end{equation}
where $a \in [0, 1]$ is the muscle activation level, $F_l$ is the
force-length relation, and $F_v$ is the force-velocity relation.
During their calculation, we need the maximum muscle stiffness
$k_{max}$, maximum muscle length $l_m$, and the maximum contractile
velocity $v_m$. We present the plots of the two relations in
Figure~\ref{fig:Fl_and_Fv}. Additional details about the setting and the
biomechanical background of the parameters can be found in
\cite{lee2006heads}.

\subsection{Skin}

Our facial skin model is automatically generated using the technique
described in \cite{lee1995realistic}. After laser-scanning the
individual's facial data, a range image and a reflectance image in
cylindrical coordinates are adapted to a well-structured face mesh.
Then the algorithm creates a dynamic skin and muscle model for the
face mesh, which contains automatically generated facial soft tissues,
estimated skull surface. Also, major muscles responsible for facial
expression are inserted to the model.

The physical simulation of the muscle-actuated facial skin model is
implemented as a discrete deformable model (DDM), where a network of
fascia nodes are connected using uni-axial springs. The force exerts
from spring $j$ on node $i$ can be written as
\begin{equation}
\vec{g}_{i}^j = c_j (l_j - l_j^r) \vec{s}_j,
\end{equation}
where $l_j$ and $l_j^r$ are the current and resting length of spring
$j$, and $\vec{s}_j=(\vec{x}_j - \vec{x}_i)/l_j$ is the spring
direction vector.

The facial skin model is also actuated by the underlying muscles. We
calculate the force exerts from muscle $j$ on node $i$ according to
the length scaling function $\Theta_1$ and the muscle-width scaling
function $\Theta_2$ as follows:
\begin{equation}
f_i^j = \Theta_1(\varepsilon_{j, i})\Theta_2(\omega_{j, i}).
\end{equation}
In \cite{lee1995realistic}, the plots of the two scaling functions, as
well as the definition of the length ratio $\varepsilon$ and muscle
width $\omega$ are explained in detail. Moreover, there are other
aspects of the generated discrete deformable model; e.g., the volume
preservation forces and skull penetration constraint forces.

\bibliographystyle{apa}
\bibliography{main}

\end{document}